\documentclass[12pt]{amsart}
\usepackage{mathrsfs}
\usepackage{amssymb} 
\usepackage{url}
\usepackage{hyperref}
\usepackage{graphicx} 
\usepackage{epstopdf}
\usepackage{tikz}
\usetikzlibrary{matrix,arrows}
\usepackage{color}
\definecolor{refcolor}{RGB}{0,0,190}
\hypersetup{
    colorlinks,
    citecolor=refcolor,
    filecolor=refcolor,
    linkcolor=refcolor,
    urlcolor=refcolor
}

\theoremstyle{definition}
\newtheorem{theorem}{Theorem}[section]
\newtheorem{definition}[theorem]{Definition}

\newcommand{\image}[3]{
\begin{center}
\begin{figure*}[htb]
\includegraphics[width=#2\textwidth]{#1.eps}
\caption{\small{\label{#1}#3}}\end{figure*}
\end{center}
}

\def\({\left(}
\def\){\right)}

\newcommand{\R}{\mathbb{R}}
\newcommand{\N}{\mathbb{N}}

\newcommand{\de}{\textnormal{d}}

\newcommand{\tn}{\textnormal}
\newcommand{\ds}{\displaystyle}

\newcommand{\ie}{\textit{i.e.} }

\newcommand{\eg}{\textit{e.g.} }
\newcommand{\etc}{\textit{etc.}}

\newcommand{\IM}{\tn{im }}

\newcommand{\mc}[1]{\mathcal{#1}}
\newcommand{\ms}[1]{\mathscr{#1}}

\newcommand{\sref}[1]{\S\ref{#1}}

\newcommand{\abs}[1]{\left|#1\right|}
\newcommand{\rank}{\textnormal{rank }}

\newcommand{\ric}{\tn{Ric}}

\newcommand{\dsfrac}[2]{\ds{\frac{#1}{#2}}}

\newcommand{\idxannih}[2]{#1{}^{#2}{}}
\newcommand{\idxcoannih}[2]{#1{}_{#2}{}}

\newcommand{\radix}[1]{\idxcoannih{#1}{\circ}}
\newcommand{\annih}[1]{\idxannih{#1}{\bullet}}
\newcommand{\coannih}[1]{\idxcoannih{#1}{\bullet}}

\newcommand{\annihg}{\coannih{g}}
\newcommand{\coannihg}{\annih{g}}
\newcommand{\metric}[1]{\langle#1\rangle}

\newcommand{\cocontr}{{{}_\bullet}}

\newcommand{\fiscal}[1]{\ms F(#1)}

\newcommand{\annihforms}[1]{\annih{\mc A}(#1)}

\newcommand{\kosz}{\mc K}
\newcommand{\der}{\nabla}
\newcommand{\dera}[1]{\der_{#1}}
\newcommand{\derb}[2]{\dera{#1}{#2}}
\newcommand{\derc}[3]{({\derb{#1}{#2}})(#3)}
\newcommand{\lder}{\der^{\flat}}
\newcommand{\ldera}[1]{\lder_{#1}}
\newcommand{\lderb}[2]{\ldera{#1}{#2}}
\newcommand{\lderc}[3]{(\lderb{#1}{#2})(#3)}

\def\hyph{-\penalty0\hskip0pt\relax}
\newcommand{\semiriem}{semi{\hyph}Riemannian}
\newcommand{\ssemiriem}{Semi{\hyph}Riemannian}
\newcommand{\semireg}{semi{\hyph}regular}
\newcommand{\ssemireg}{Semi{\hyph}regular}
\newcommand{\quasireg}{quasi{\hyph}regular}

\newcommand{\nondeg}{non{\hyph}degenerate}

\newcommand{\flrw}{Friedmann-Lema\^itre-Robertson-Walker}
\newcommand{\FLRW}{FLRW}
\newcommand{\schw}{Schwarzschild}
\newcommand{\rn}{Reissner-Nordstr\"om}
\newcommand{\kn}{Kerr-Newman}
\newcommand{\hor}{Ho{\v{r}}ava}
\newcommand{\HL}{\hor-Lifschitz}

\newcommand{\CS}{\mathbb{S}}
\newcommand{\CT}{\mathbb{T}}
\newcommand{\CM}{\mathbb{M}}

\begin{document} 

\title{The Geometry of Black Hole singularities}

\author{Ovidiu Cristinel \ Stoica}
\date{\today. Horia Hulubei National Institute for Physics and Nuclear Engineering, Bucharest, Romania. E-mail: holotronix@gmail.com}

\begin{abstract}
Recent results show that important singularities in General Relativity can be naturally described in terms of finite and invariant canonical geometric objects.
Consequently, one can write field equations which are equivalent to Einstein's at non-singular points, but in addition remain well-defined and smooth at singularities. 

The black hole singularities appear to be less undesirable than it was thought, especially after we remove the part of the singularity due to the coordinate system. Black hole singularities are then compatible with global hyperbolicity, and don't make the evolution equations break down, when these are expressed in terms of the appropriate variables.

The charged black holes turn out to have smooth potential and electromagnetic fields in the new atlas. Classical charged particles can be modeled, in General Relativity, as charged black hole solutions. Since black hole singularities are accompanied by dimensional reduction, this should affect Feynman's path integrals. Therefore, it is expected that singularities induce dimensional reduction effects in Quantum Gravity. These dimensional reduction effects are very similar to those postulated in some approaches to making Quantum Gravity perturbatively renormalizable. This may provide a way to test indirectly the effects of singularities, otherwise inaccessible.
\keywords{singular semi-Riemannian geometry,general relativity,singularities,quantum gravity,dimensional reduction.}
\end{abstract}


\maketitle

\setcounter{tocdepth}{1}
\tableofcontents

\section{Introduction}

For millennia, space was considered the fixed background -- the arena where physical phenomena took place. Special Relativity changed this, by proposing spacetime as the new arena. Then, while trying to extend the success of Special Relativity to non-inertial frames and gravity, Einstein realized that one should let go the idea of an immutable background, and General Relativity (GR) was born. There is a very deep interdependence between matter and the geometry of spacetime, encoded in Einstein's equation. Its predictions were tested with high accuracy, and confirmed.

However, the task of decoding the way our universe works from something as abstract as Einstein's equation is not easy, and we are far from grasping all of its consequences.
For instance, even from the beginning, when {\schw} proposed his model for the exterior of a spherically symmetric object, Einstein's equations led to infinities \cite{Scw16a,Scw16b}. The {\schw} metric tensor becomes infinite at $r=0$ and on the event horizon -- where $r=2m$. The big bang also exhibited a singularity \cite{FRI22de,FRI99en,FRI24,LEM27,ROB35I,ROB35II,ROB35III,WAL37}.

The first reaction to the singularities was to somehow minimize their importance, on the grounds that they are exceptions due to the perfect symmetry of the solutions. This hope was ruined by the theorems of Penrose \cite{Pen65,Pen69} and Hawking \cite{Haw66i,Haw66ii,Haw67iii,HP70}, showing that the singularities are predicted to occur in GR under very general conditions, and are not caused by the perfect symmetry.

Singularities, hidden by the event horizon or naked, are very well researched in the literature (for example \cite{Pen69,Pen78,Pen79,isham1981qg2,Pen98}, \cite{barve1997naked,barve1998particle,barve1998quantum,barve2000spherical}, \cite{joshi2013spacetime}, and references therein).

Interesting results concerning singularities were obtained in some modified gravity theories, e.g. $f(R)$ gravity (\cite{buchdahl1970fR,starobinsky1980fR,borisov2012fR,hendi2011black,olmo2011palatini-fR} and references therein). Another way to avoid singularities was proposed in non-linear electrodynamics \cite{corda2010removingBHsingularities}. 

In addition to the singularities, infinities occur in GR when we try to quantize gravity, because gravity is perturbatively nonrenormalizable \cite{HV74qg,GS86uvgr}. It is expected by many that a solution to the problem of quantization will also remove the singularities. For example, {\em Loop quantum cosmology} obtained significant positive results in showing that quantum effects may prevent the occurrence of singularities \cite{bojowald2003absenceLQC,ashtekar2011LQC,visinescu2009bianchi,visinescu2012bianchi}.

There is another possibility: the problem of singularities may be in fact not due to GR, but to our limited understanding of GR. Therefore, it would be useful to better understand singularities, even in the eventuality that a better theory will replace GR. In the following we review some recent results showing that by confronting singularities, we realize that they are not that undesirable \cite{Sto13a}. Moreover, new possibilities open also for the Quantum Gravity problem.

\section{The problem of singularities in General Relativity}

\subsection{Two types of singularities}

Not all singularities are born equal. We can roughly classify the singularities in two types:

\begin{enumerate}
	\item {\em Malign singularities}: some of the components of the metric are divergent: $g_{ab}\to\infty$. 
  \item {\em Benign singularities}: $g_{ab}$ are smooth and finite, but $\det g\to 0$.
\end{enumerate}

Benign singularities turn out to be, in many cases, manageable \cite{Sto11a,Sto11b,Sto12b}. The infinities simply disappear, if we use different geometric objects to write the equations and describe the phenomena. At points where the metric is {\nondeg}, the proposed description is equivalent to the standard one. But, in addition, it works also at the points where the metric becomes degenerate.

Malign singularities appear in the black hole solutions. They appear to be malign because the coordinates in which are represented are singular. In non-singular coordinates, they become benign \cite{Sto11e,Sto11f,Sto11g}. This is somewhat similar to the case of the apparent singularity on the event horizon, which turned out to be a coordinate singularity, and not a genuine one \cite{eddington1924comparison,finkelstein1958past}.

\subsection{What is wrong with singularities}

The geometry of spacetime is encoded in the metric tensor. To write down field equations, we have to use partial derivatives. In curved spaces, partial derivatives are replaced by {\em covariant derivatives}. They are defined with the help of the {\em Levi-Civita connection}, which takes into account the parallel translations, to compare fields at infinitesimally closed points. The covariant derivative is written using the Christoffel symbol of the second kind, obtained from the metric tensor by
\begin{equation}
\label{eq_christoffel_second_kind}
		\Gamma^{c}{}_{ab} = \ds{\frac 1 2} {g^{cs}}(
		\partial_a g_{bs} + \partial_b g_{sa} - \partial_s g_{ab}).
\end{equation}
It can be used to define the Riemann curvature tensor
\begin{equation}
\label{eq_riemann_curvature}
		R^{d}{}_{abc} = \Gamma^{d}{}_{ac,b} - \Gamma^{d}{}_{ab,c} + \Gamma^{d}{}_{bs}\Gamma^{s}{}_{ac} - \Gamma^{d}{}_{cs}\Gamma^{s}{}_{ab}.
\end{equation}
It plays a major part in the Einstein equation
\begin{equation}
\label{eq_einstein_idx}
	G_{ab} + \Lambda g_{ab} = \kappa T_{ab},
\end{equation}
since
\begin{equation*}
	G_{ab} = R_{ab} - \frac 1 2 R g_{ab},
\end{equation*}
where $R_{ab} = R^s{}_{asb}$ is the Ricci tensor, 
and $R = R^s{}_{s}$ is the scalar curvature.

In the case of malign singularities, since some of metric's components are singular, the geometric objects like the Levi-Civita connection and the Riemann curvature tensor are singular too. Therefore, it seems that the situation of malign singularities is hopeless.

Even in the case of benign singularities, when the metric is smooth, but its determinant $\det g\to 0$, the usual Riemannian objects are singular. For example, the covariant derivative can't be defined, because the inverse of the metric, $g^{ab}$, becomes singular ($g^{ab}\to\infty$ when $\det g\to 0$). This makes the Christoffel's symbols of the second kind \eqref{eq_christoffel_second_kind}, and the Riemann curvature \eqref{eq_riemann_curvature} singular.

It is therefore understandable why singularities were considered unsolvable problems for so many years.

\subsection{From singular to non-singular -- a dictionary}

The main variables which appear in the equations are indeed singular. But we can replace them with new variables, which are equivalent to the original ones on the domain where both are defined. Sometimes, we can choose the new variables so that the equations remain valid at points where the original ones were singular.

The geometric objects of interest that become singular when the metric is degenerate are the Levi-Civita connection \eqref{eq_christoffel_second_kind}, the Riemann curvature \eqref{eq_riemann_curvature}, the Ricci and the scalar curvatures. If the metric is {\nondeg}, the Christoffel symbols of the first kind are equivalent to those of the second kind, in the sense that by knowing one of them, we can obtain the other one. Similarly, the Riemann curvature $R^{a}_{bcd}$ is equivalent to $R_{abcd}$, the Ricci and scalar curvatures are equivalent to their densitized versions and to their Kulkarni-Nomizu products (see equation \ref{eq_kulkarni_nomizu}) with the metric. 
In some important cases, these equivalent objects remain non-singular even when the metric is degenerate, \cite{Sto11a,Sto12b}. We summarize these cases in Table \ref{table_objects}.

\begin{table}[htdp]
	\begin{center}
		 \begin{tabular}{ l | l | l}
			 \hline
			 Singular & Non-Singular & When g is... \\ \hline \hline
			 $\Gamma^c{}_{ab}$ (2-nd) & $\Gamma_{abc}$ (1-st) & smooth \\ \hline
			 $R^d{}_{abc}$ & $R_{abcd}$ & {\semireg} \\ \hline
			 $R_{ab}$ & $R_{ab}\sqrt{\abs{\det g}}^W,\,W\leq 2$ & {\semireg} \\ \hline
			 $R$ & $R\sqrt{\abs{\det g}}^W,\,W\leq 2$ & {\semireg} \\ \hline
			 $\ric$ & $\ric \circ g$ & {\quasireg} \\ \hline
			 $R$ & $R g \circ g$ & {\quasireg} \\ \hline
			 \hline
		 \end{tabular}
	\end{center}
	\caption{Singular objects and their non-singular equivalents.}
	\label{table_objects}
\end{table}

\section{The mathematical methods: Singular {\ssemiriem} Geometry}
\label{s_ssrg}

\subsection{Singular {\ssemiriem} Geometry}

We review the main mathematical tool on which the results presented here are based, named {\em Singular {\ssemiriem} Geometry} \cite{Sto11a,Sto11b}.
Singular {\ssemiriem} Geometry is mainly concerned with the study of singular {\semiriem} manifolds.

\begin{definition}(see \cite{Sto11a}, \cite{Kup87b})
\label{def_sing_semiRiemm_man}
A \textit{singular {\semiriem} manifold} $(M,g)$ consists in a differentiable manifold $M$, and a symmetric bilinear form $g$ on $M$, named \textit{metric tensor} or \textit{metric}.
\end{definition}

If $g$ is {\nondeg}, then $(M,g)$ is just a {\em{\semiriem} manifold}. If in addition $g$ is positive definite, $(M,g)$ is named \textit{Riemannian manifold}. In General Relativity {\semiriem} manifolds are normally used, but when we are dealing with singularities, it is natural to use the Singular {\ssemiriem} Geometry, which is more general.

\subsection{Properties of the degenerate inner product}

Let $(V,g)$ be an inner product vector space. Let $\flat:V\to V^*$ be the morphism defined by $u\mapsto \annih{u}:=\flat(u)=u^\flat=g(u,\_)$. We define the {\em radical} of $V$ as the set of isotropic vectors in $V$, $\radix{V}:=\ker\flat=V^\perp$. We define the {\em radical annihilator} space of $V$ as the image of $\flat$, $\annih{V}:=\IM{\flat}\subset V^*$. The inner product $g$ induces on $\annih{V}$ an inner product, defined by $\annihg(u_1^\flat,u_1^\flat):=g(u_1,u_2)$. This one is the inverse of $g$ if and only if $\det g\neq 0$. The {\em coannihilator} is the quotient space $\coannih{V}:=V/\radix{V}$, given by the equivalence classes of the form $u+\radix{V}$. On the coannihilator $\coannih{V}$, the metric $g$ induces an inner product $\coannihg(u_1+\radix{V},u_2+\radix{V}):=g(u_1,u_2)$.

Let $p\in M$. In the following, we will denote by $\radix{T}_p M\leq T_p M$ the radical of the tangent space at $p$, by $\annih{T}_p M\leq T^*_p M$ the radical annihilator, and by $\coannih{T}_p M$ the coannihilator.

We have seen that one important problem which appears when the metric becomes degenerate is that it doesn't admit an inverse $g^{ab}$, and fundamental tensor operations like raising indices and contractions between covariant indices are no longer defined. But we can use the reciprocal metric $\annihg$ to define metric contraction between covariant indices, for tensors that live in tensor products between $T_p M$ and the subspace $\annih{T}_p M$. This turned out to be enough for some important singularities in General Relativity.

\subsection{Covariant derivative}

Because at points where the metric is degenerate there is no inverse metric, the Levi-Civita connection is not defined. Then, how can we derivate? We will see that in some cases, which turn out to be enough for our purposes, we still can derivate.

\subsubsection{The Koszul object}

Let $X,Y,Z$ be vector fields on $M$. We define the {\em Koszul object} as
\begin{equation}
\label{eq_Koszul_form}
\begin{array}{llll}
	\kosz(X,Y,Z) &:=&\ds{\frac 1 2} \{ X \metric{Y,Z} + Y \metric{Z,X} - Z \metric{X,Y} \\
	&&\ - \metric{X,[Y,Z]} + \metric{Y, [Z,X]} + \metric{Z, [X,Y]}\}.
\end{array}
\end{equation}

Its components in local coordinates are just Christoffel's symbols of the first kind:
\begin{equation}
\label{eq_Koszul_form_coord}
	\kosz_{abc}=\kosz(\partial_a,\partial_b,\partial_c)=\ds{\frac 1 2} (
	\partial_a g_{bc} + \partial_b g_{ca} - \partial_c g_{ab}) = \Gamma_{abc},
\end{equation}

If the metric is {\nondeg}, one defines the Levi-Civita connection uniquely, by raising an index of the Koszul object:
\begin{equation}
\label{eq_koszul_formula_inv}
	\derb X Y = \kosz(X,Y,\_)^{\sharp}.
\end{equation}

But if the metric is degenerate, one cannot raise the index, and we will have to avoid the usage of the Levi-Civita connection. Luckily, we can do what we do with the Levi-Civita connection and more, just by using the Koszul object instead.

\subsubsection{The covariant derivatives}
We define the {\em lower covariant derivative} of a vector field $Y$ in the direction of a vector field $X$ by
\begin{equation}
\label{eq_l_cov_der_vect}
\lderc XYZ := \kosz(X,Y,Z).
\end{equation}
This is not quite a true covariant derivative, because it doesn't map vector fields to vector fields, but to $1$-forms. However, we can use it to replace the covariant derivative of vector fields, and it is equivalent to it if the metric is {\nondeg}.

If the Koszul object satisfies the condition that $\kosz(X,Y,W)=0$ for any $W\in\Gamma(\radix{T}M)$, then the singular {\semiriem} manifold $(M,g)$ is named {\em radical stationary}. In this case, it makes sense to contract in the third slot of the Koszul object, and define by this covariant derivatives of differential forms.
The covariant derivative of differential forms is defined by
\begin{equation*}
	\left(\der_X\omega\right)(Y) := X\left(\omega(Y)\right) - \annihg(\lderb X Y,\omega),
\end{equation*}
if $\omega\in \annihforms{M}:=\Gamma(\annih{T}M)$.
More general,

\begin{equation*}
	\der_X(\omega_1\otimes\ldots\otimes\omega_s) := \der_X(\omega_1)\otimes\ldots\otimes\omega_s +\ldots + \omega_1\otimes\ldots\otimes\der_X(\omega_s).
\end{equation*}

The covariant derivative of a tensor $T\in\Gamma(\otimes^k_M\annih{T}M)$ is defined as
\begin{equation*}
\begin{array}{lll}
	\left(\nabla_X T\right)(Y_1,\ldots,Y_k) &=& X\left(T(Y_1,\ldots,Y_k)\right) \\
	&&- \sum_{i=1}^k\kosz(X,Y_i,\cocontr)T(Y_1,,\ldots,\cocontr,\ldots,Y_k).
\end{array}
\end{equation*}

\subsection{Riemann curvature tensor. {\ssemireg} manifolds.}

Let $(M,g)$ be a radical stationary manifold. Then, the {\em Riemann curvature tensor} is defined as
\begin{equation}
\label{eq_riemann_curvature_explicit}
	R(X,Y,Z,T) = \derc X {{\ldera Y}Z}T - \derc Y {{\ldera X}Z}T - \lderc {[X,Y]}ZT.
\end{equation}

The components of the Riemann curvature tensor in local coordinates are
\begin{equation}
	R_{abcd}= \partial_a \kosz_{bcd} - \partial_b \kosz_{acd} + (\kosz_{ac\cocontr}\kosz_{bd\cocontr} - \kosz_{bc\cocontr}\kosz_{ad\cocontr}).
\end{equation}

The Riemann curvature tensor has the same symmetry properties as in Riemannian geometry, and is radical-annihilator in each of its slots.

A singular {\semiriem} manifold is called {\em \semireg} \cite{Sto11a} if:
\begin{equation}
	\dera X {\ldera Y}Z \in \annihforms M.
\end{equation}
An equivalent condition is
\begin{equation}
	\kosz(X,Y,\cocontr)\kosz(Z,T,\cocontr) \in \fiscal M.
\end{equation}
It is easy to see that the Riemann curvature of {\semireg} manifolds is smooth.

\subsection{Examples of {\semireg} {\semiriem} manifolds}
\label{s_semi_reg_semi_riem_man_example}

We present some examples of {\semiriem} manifolds \cite{Sto11a,Sto11b}.

\subsubsection{Isotropic singularities}

Isotropic singularities have the form
$$g=\Omega^2\tilde g,$$
where $\tilde g$ is a {\nondeg} bilinear form on $M$.

Such singularities were studied in connection to some cosmological models \cite{Tod87,Tod90,Tod91,Tod92,CN98,AT99i,AT99ii,Tod02,Tod03}.

\subsubsection{Degenerate warped products}

Warped products are products of two {\semiriem} manifolds $(B,g_B)$ and $(F,g_F)$, so that the metric on the manifold $F$ is scaled by a scalar function $f$ defined on the  manifold $B$  \cite{ONe83}. The warped product has the form:
\begin{equation}
\label{eq_wp_metric}
	\de s^2=\de s_B^2 + f^2(p)\de s_F^2.
\end{equation}
Normally, the warping function $f$ is taken to be strictly positive at all points of $B$. However, it may happen to vanish at some points, and in this case the result is a singular {\semiriem} manifold. The resulting manifold is {\semireg} \cite{Sto11b}. Moreover, if the manifolds $B$ and $F$ are radical stationary, and if $\de f\in  \annihforms M$, their warped product is radical stationary. If $B$ and $F$ are {\semireg}, $\de f\in \annihforms M$, and $\nabla_X\de f\in  \annihforms M$ for any vector field $X$, then $B\times_f F$ is {\semireg} \cite{Sto11b}.

\section{Einstein equations at singularities}
\label{s_einstein_tensor_densitized}

We discuss now two equations which are equivalent to Einstein's when the metric is {\nondeg}, but remain smooth and finite also at some singularities. The first equation remains smooth at {\semireg} singularities, while the second at {\quasireg} singularities.

\subsection{Einstein's equation on {\semireg} spacetimes}

\subsubsection{The densitized Einstein equation}

Consider the following densitized version of the Einstein equation
\begin{equation}
\label{eq_einstein:densitized}
	G\det g + \Lambda g\det g = \kappa T\det g,
\end{equation}
or, in coordinates or local frames,
\begin{equation}
\label{eq_einstein_idx:densitized}
	G_{ab}\det g + \Lambda g_{ab}\det g = \kappa T_{ab}\det g.
\end{equation}
If the metric is {\nondeg}, this equation is equivalent to the Einstein equation, the only difference being the factor $\det g\neq 0$.
But what happens if the metric becomes degenerate? In this case, it is not allowed to divide by $\det g$, because this is $0$.

On four-dimensional {\semireg} spacetimes Einstein tensor density $G\det g$ is smooth \cite{Sto11a}.
Hence,  the proposed densitized Einstein equation \eqref{eq_einstein:densitized} is smooth, and non-singular.
If the metric is regular, this equation is equivalent to the Einstein equation.

\subsubsection{{\FLRW} spacetimes}
\label{s_semireg_examples_flrw}

To better understand black hole singularities, which will be discussed later, we start by taking a look at the {\flrw} (\FLRW) singularities, which are benign. Black hole singularities are malign, but can be made benign by removing the coordinate singularity (see sections \sref{s_schw}, \sref{s_rn}, and \sref{s_kn}).

{\FLRW} spacetimes are examples of degenerate warped products, with the metric defined by
\begin{equation}
	\de s^2 = -\de t^2 + a^2(t)\de\Sigma^2,
\end{equation}
where
\begin{equation}
\label{eq_flrw_sigma_metric}
\de\Sigma^2 = \dsfrac{\de r^2}{1-k r^2} + r^2\(\de\theta^2 + \sin^2\theta\de\phi^2\),
\end{equation}
where $k=1$ for $S^3$, $k=0$ for $\R^3$, and $k=-1$ for $H^3$. It follows that they are {\semireg}.

Since the {\FLRW} singularities are warped products, they are {\semireg}. Therefore, we can expect that the densitized Einstein equation holds. In fact, in \cite{Sto11h} is shown more than that, as we will see now.

The {\FLRW} stress-energy tensor is
\begin{equation}
\label{eq_friedmann_stress_energy}
T^{ab} = \(\rho+p\)u^a u^b + p g^{ab},
\end{equation}
where $u^a$ is the timelike vector field $\partial_t$, normalized.
The scalar $\rho$ represents the mass density, and $p$ the pressure density. From the stress-energy tensor \eqref{eq_friedmann_stress_energy}, in the case of a homogeneous and isotopic universe, follow the {\em Friedmann equation}
\begin{equation}
\label{eq_friedmann_density}
\rho = \dsfrac{3}{\kappa}\dsfrac{\dot{a}^2 + k}{a^2},
\end{equation}
and the {\em acceleration equation}
\begin{equation}
\label{eq_acceleration}
\rho + 3p = -\dsfrac{6}{\kappa}\dsfrac{\ddot{a}}{a}.
\end{equation}

Equations \eqref{eq_friedmann_density} and \eqref{eq_acceleration} show that the scalars $\rho$ and $p$ are singular for $a=0$. But $\rho$ and $p$ represent the mass and pressure densities the orthonormal frame obtained by normalizing the comoving frame $(\partial_t,\partial_x,\partial_y,\partial_z)$, where $(x,y,z)$ are coordinates on the space manifold $S$. The mass and pressure density can be identified with the scalars $\rho$ and $p$ only in an orthogonal frame. But at the singularity $a=0$ there is no orthonormal frame, so we should not normalize the comoving frame. In the general, non-normalized case, the actual densities contain in fact the factor $\sqrt{-g}(=a^3 \sqrt{g_{\Sigma}})$,
\begin{equation}
\label{eq_substitution_densities}
\begin{array}{l}
\bigg\{
\begin{array}{ll}
	\widetilde\rho = \rho \sqrt{-g} = \rho a^3 \sqrt{g_{\Sigma}} \\
	\widetilde p = p \sqrt{-g} = p a^3 \sqrt{g_{\Sigma}}. \\
\end{array}
\\
\end{array}
\end{equation}

The Friedmann and the acceleration equations become
\begin{equation}
\label{eq_friedmann_density_tilde}
\widetilde\rho = \dsfrac{3}{\kappa}a\(\dot a^2 + k\) \sqrt{g_{\Sigma}},
\end{equation}
and
\begin{equation}
\label{eq_acceleration_tilde}
\widetilde\rho + 3\widetilde p = -\dsfrac{6}{\kappa}a^2\ddot{a} \sqrt{g_{\Sigma}}.
\end{equation}

We see that $\widetilde \rho$ and $\widetilde p$ are smooth, and so is the densitized stress-energy tensor
\begin{equation}
T_{ab}\sqrt{-g} = \(\widetilde\rho+\widetilde p\)u_a u_b + \widetilde p g_{ab}.
\end{equation}
We obtain a densitized Einstein equation, from which equation \eqref{eq_einstein:densitized} follows by multiplying with $\sqrt{-g}$.

Hence, the {\FLRW} solution is described by smooth densities even at the big bang singularity. Moreover, the solution extends beyond the singularity.

\subsection{Einstein's equation on {\quasireg} spacetimes}

\subsubsection{The Ricci decomposition}

Let $(M,g)$ be an $n$-dimensional {\semiriem} manifold. The Riemann curvature decomposes algebraically \cite{ST69,BESS87,GHLF04} as
\begin{equation}
\label{eq_ricci_decomposition}
	R_{abcd} = S_{abcd} + E_{abcd} + C_{abcd},
\end{equation}
where 
\begin{equation}
	S_{abcd} = \dsfrac{1}{n(n-1)}R(g\circ g)_{abcd},
\end{equation}
\begin{equation}
	E_{abcd} = \dsfrac{1}{n-2}(S \circ g)_{abcd},
\end{equation}
\begin{equation}
\label{eq_ricci_traceless}
S_{ab} := R_{ab} - \dsfrac{1}{n}Rg_{ab},
\end{equation}
where $\circ$ denotes the Kulkarni-Nomizu product:
\begin{equation}
\label{eq_kulkarni_nomizu}
	(h\circ k)_{abcd} := h_{ac}k_{bd} - h_{ad}k_{bc} + h_{bd}k_{ac} - h_{bc}k_{ad}.
\end{equation}

If the Riemann curvature tensor on a {\semireg} manifold $(M,g)$ admits such a decomposition so that all of its terms are smooth, $(M,g)$ is said to be {\em \quasireg}.

\subsubsection{The expanded Einstein equation}

For dimension $n=4$, in \cite{Sto12b} we introduced the {\em expanded Einstein equation}
\begin{equation}
\label{eq_einstein_expanded}
	(G\circ g)_{abcd} + \Lambda (g\circ g)_{abcd} = \kappa (T\circ g)_{abcd}
\end{equation}

or, equivalently,
\begin{equation}
\label{eq_einstein_expanded_explicit}
	2 E_{abcd} - 6 S_{abcd} + \Lambda (g\circ g)_{abcd} = \kappa (T\circ g)_{abcd}.
\end{equation}

It is equivalent to Einstein's equation if the metric is {\nondeg}, but in addition extends smoothly at {\quasireg} singularities.

\subsubsection{Examples of {\quasireg} singularities}

As shown in \cite{Sto12b}, the following are examples of {\quasireg} singularities:
\begin{itemize}
	\item 
	Isotropic singularities.
	\item 
	Degenerate warped products $B\times_f F$ with $\dim B=1$ and $\dim F=3$.
	\item 
	{\FLRW} singularities, as a particular case of degenerate warped products \cite{Sto12a}.
	\item 
	{\schw} singularities (after removing the coordinates singularity, see section \sref{s_schw}). The question whether the {\rn} and {\kn} singularities are {\quasireg}, or at least {\semireg}, is still open.
\end{itemize}

\subsubsection{The Weyl curvature hypothesis and {\quasireg} singularities}

To explain the low entropy at the big bang and the high homogeneity of the universe, Penrose emitted the Weyl curvature hypothesis, stating that the Weyl curvature tensor vanishes at the big bang singularity \cite{Pen79}.

From equation \eqref{eq_ricci_decomposition}, the {\em Weyl curvature tensor} is
\begin{equation}
\label{eq_weyl_curvature}
	C_{abcd} = R_{abcd} - S_{abcd} - E_{abcd}.
\end{equation}

In \cite{Sto12c} it was shown that, when approaching a {\quasireg} singularity, $C_{abcd} \to 0$ smoothly.
Because of this, any {\quasireg} big bang satisfies the Weyl curvature hypothesis. In \cite{Sto12c}  it has also been shown that a very large class of big bang singularities, which are not homogeneous or isotropic, are {\quasireg}.

\subsection{Taming a malign singularity}

We have seen that when the singularity is benign, {\ie} the singularity is due to the degeneracy of the metric tensor, which is smooth, there are important cases when we can obtain a complete description of the fields and their evolution, in terms of finite quantities.

But what can we do if the singularities are malign? This case is important, since all black hole singularities are malign.
In \cite{Sto11e,Sto11f,Sto11g} we show that, although the black hole singularities appear to be malign, we can make them benign, by a proper choice of coordinates. This is somewhat analog to the method used in \cite{eddington1924comparison} and \cite{finkelstein1958past} to show that the event horizon singularity is not a true singularity, being due to coordinates. In the following sections, we will review these results.

\section{{\schw} singularity is {\semireg}}
\label{s_schw}

The {\schw} metric is given in {\schw} coordinates by
\begin{equation}
\label{eq_schw_schw}
\de s^2 = -\(1-\dsfrac{2m}{r}\)\de t^2 + \(1-\dsfrac{2m}{r}\)^{-1}\de r^2 + r^2\de\sigma^2,
\end{equation}
where
\begin{equation}
\label{eq_sphere}
\de\sigma^2 = \de\theta^2 + \sin^2\theta \de \phi^2.
\end{equation}

Let's change the coordinates to
\begin{equation}
\label{eq_coordinate_semireg}
\begin{array}{l}
\bigg\{
\begin{array}{ll}
r &= \tau^2 \\
t &= \xi\tau^4. \\
\end{array}
\\
\end{array}
\end{equation}

The four-metric becomes
\begin{equation}
\label{eq_schw_analytic_tau_xi}
\de s^2 = -\dsfrac{4\tau^4}{2m-\tau^2}\de \tau^2 + (2m-\tau^2)\tau^{4}\(4\xi\de\tau + \tau\de\xi\)^2 + \tau^4\de\sigma^2,
\end{equation}
which is analytic and {\semireg}  at $r=0$ \cite{Sto11e}.

The problems were fixed by a coordinate change. Doesn't this mean that the singularity depends on the coordinates? Well, this deserves an explanation. Changing the coordinates doesn't make a singularity appear or disappear, if the coordinate transformation is a local diffeomorphism. But a regular tensor can become singular, or a singular tensor can become regular, if the coordinate transformation itself is singular. This situation is very similar to that of the event horizon singularity $r=2m$ of the {\schw} metric, in {\schw} coordinates \eqref{eq_schw_schw}. This singularity vanishes when we go to the Eddington-Finkelstein coordinates. This proves that the Eddington-Finkelstein coordinates are from the correct atlas, while the original {\schw} coordinates were in fact singular at $r=2m$. In our case, the coordinate transformation \eqref{eq_coordinate_semireg} allows us to move to an atlas in which the metric is analytic and {\semireg}, showing that the {\schw} coordinates were in fact singular at $r=0$.

\section{Charged and non-rotating black holes}
\label{s_rn}

Charged non-rotating black holes are described by the {\rn} metric,
\begin{equation}
\label{eq_rn_metric}
\de s^2 = -\left(1-\dsfrac{2m}{r} + \dsfrac{q^2}{r^2}\right)\de t^2 + \left(1-\dsfrac{2m}{r} + \dsfrac{q^2}{r^2}\right)^{-1}\de r^2 + r^2\de\sigma^2,
\end{equation}

To make the singularity benign, we choose the new coordinates $\rho$ and $\tau$ \cite{Sto11f}, so that
\begin{equation}
	\begin{array}{l}
	\bigg\{
	\begin{array}{ll}
	t &= \tau\rho^T \\
	r &= \rho^S \\
	\end{array}
	\\
	\end{array}
\end{equation}

In the new coordinates, the metric has the following form
\begin{equation}
\label{eq_rn_ext_ext}
\de s^2 = - \Delta\rho^{2T-2S-2}\left(\rho\de\tau + T\tau\de\rho\right)^2 + \dsfrac{S^2}{\Delta}\rho^{4S-2}\de\rho^2 + \rho^{2S}\de\sigma^2,
\end{equation}
\begin{equation}
	\tn{where \,\,}\Delta := \rho^{2S} - 2m \rho^{S} + q^2.
\end{equation}

To remove the infinity of the metric at $r=0$ and ensure analiticity, we have to choose
\begin{equation}
	\begin{array}{l}
	\bigg\{
	\begin{array}{ll}
	S \geq 1 \\
	T \geq S + 1.
	\end{array}
	\\
	\end{array}
\end{equation}

In the {\rn} coordinates $(t,r,\phi,\theta)$, the electromagnetic potential is singular at $r=0$,
\begin{equation}
A = -\dsfrac q r \de t.
\end{equation}
But in the new coordinates $(\tau,\rho,\phi,\theta)$, the electromagnetic potential is
\begin{equation}
A = -q\rho^{T-S-1}\left(\rho\de\tau + T\tau\de\rho\right),
\end{equation}
the electromagnetic field is
\begin{equation}
F = q(2T-S)\rho^{T-S-1}\de\tau \wedge\de\rho,
\end{equation}
and they are analytic everywhere, including at the singularity $\rho=0$ \cite{Sto11f}.

The proposed coordinates define a space+time foliation only if $T\geq 3S$ \cite{Sto11f}.

\section{Rotating black holes}
\label{s_kn}

Electrically neutral rotating black holes are represented by the Kerr solution. If they are also charged, they are described by the very similar {\kn} solution.

Consider the space $\R\times\R^3$, where $\R$ represents the time coordinate, and $\R^3$ the space, parameterized by the spherical coordinates $(r,\phi,\theta)$. The rotation is characterized by the parameter $a\geq 0$, $m\geq 0$ is the mass, and $q\in\R$ the charge. The following notations are useful
\begin{equation*}
\Sigma(r,\theta) := r^2 + a^2 \cos^2 \theta,\,
\end{equation*}
\begin{equation*}
\Delta(r) := r^2 - 2 m r + a^2 + q^2.
\end{equation*}
The non-vanishing components of the {\kn} metric are \cite{Wal84}
\begin{equation*}
g_{tt} = - \frac{\Delta(r) - a^2\sin^2\theta}{\Sigma(r,\theta)},\,	
\end{equation*}
\begin{equation*}
g_{rr} = \frac{\Sigma(r,\theta)}{\Delta(r)},\,
\end{equation*}
\begin{equation*}
g_{\theta\theta} = \Sigma(r,\theta),
\end{equation*}
\begin{equation*}
g_{\phi\phi} = \frac{(r^2 + a^2)^2 -\Delta(r) a^2\sin^2\theta}{\Sigma(r,\theta)}\sin^2\theta,\,
\end{equation*}
\begin{equation*}
g_{t\phi} = g_{\phi t} = - \frac{2a\sin^2\theta(r^2 + a^2 - \Delta(r))}{\Sigma(r,\theta)}.
\end{equation*}

In \cite{Sto11g} it was shown that in the coordinates $\tau$, $\rho$, and $\mu$, defined by
\begin{equation}
\label{eq_coordinate_ext_ext_kn}
\begin{array}{l}
\left\{
\begin{array}{ll}
t &= \tau\rho^{\CT}, \\
r &= \rho^{\CS}, \\
\phi &= \mu\rho^{\CM}, \\
\theta &= \theta, \\
\end{array}
\right.
\\
\end{array}
\end{equation}
where ${\CS},{\CT},{\CM}\in\N$ are positive integers so that
\begin{equation}
\label{eq_metric_smooth_cond_kn}
\begin{array}{l}
\left\{
\begin{array}{ll}
{\CS} &\geq 1 \\
{\CT} &\geq {\CS} + 1 \\
{\CM} &\geq {\CS} + 1,
\end{array}
\right.
\\
\end{array}
\end{equation}
the metric is analytic.

Not only the metric becomes analytic in the proposed coordinates, but also the electromagnetic potential and electromagnetic field.
The electromagnetic potential of the {\kn} solution is, in the standard coordinates, the $1$-form
\begin{equation}
\label{eq_kn_electromagnetic_potential}
A = -\dsfrac{qr}{\Sigma(r,\theta)}(\de t - a\sin^2\theta\de\phi).
\end{equation}
In the proposed coordinates is 
\begin{equation}
\label{eq_kn_electromagnetic_potential_smooth}
A = -\dsfrac{q\rho^{\CS}}{\Sigma(r,\theta)}(\rho^{\CT}\de\tau + {\CT}\tau\rho^{{\CT}-1}\de\rho - a\sin^2\theta\rho^{\CM}\de\mu).
\end{equation}
which is smooth \cite{Sto11g}.
The electromagnetic field $F = \de A$ is smooth too.

\section{Global hyperbolicity and information loss}

\subsection{Foliations with Cauchy hypersurfaces}

While Einstein's equation describes the relation between geometry and matter in a block-world view of the universe, there are equivalent formulations which express this relation from the perspective of the time evolution. Einstein's equation can be expressed in terms of a Cauchy problem \cite{ChoquetBruhat1952CauchyGR,ADM62,ACY00,CnY02,rodnianski2006cauchy,senovilla1998singularity}.

The standard black hole solutions pose two main problems to the Cauchy problem. First, the solutions have malign singularities. Second, they have in general Cauchy horizons.
Luckily, there's more than one way to skin a black hole.

The evolution equations make sense at least locally, if the singularities are benign. The black hole singularities appear to be malign in the coordinates used so far, but by removing the coordinate's contribution to the singularity, they become benign.
Even so, to formulate initial value problems globally, spacetime has to admit space+time foliations.
The spacelike hypersurfaces have to be Cauchy surfaces, in other words, the global hyperbolicity condition has to be true. The topology of the spacelike hypersurfaces must remain independent on the time $t$, although the metric is allowed to become degenerate.
This seems to be prevented in the case of {\rn} and {\kn} black holes, by the existence of Cauchy horizons.
As shown in \cite{Sto12e}, the stationary black hole singularities admit such foliations, and are therefore compatible with the condition of global hyperbolicity.

\subsection{{\schw} black holes}
\label{s_schw_foliation}

In the proposed coordinates for the {\schw} black hole, the metric extends analytically beyond the $r=0$ singularity (fig. \ref{semireg-schwarzschild}).

\image{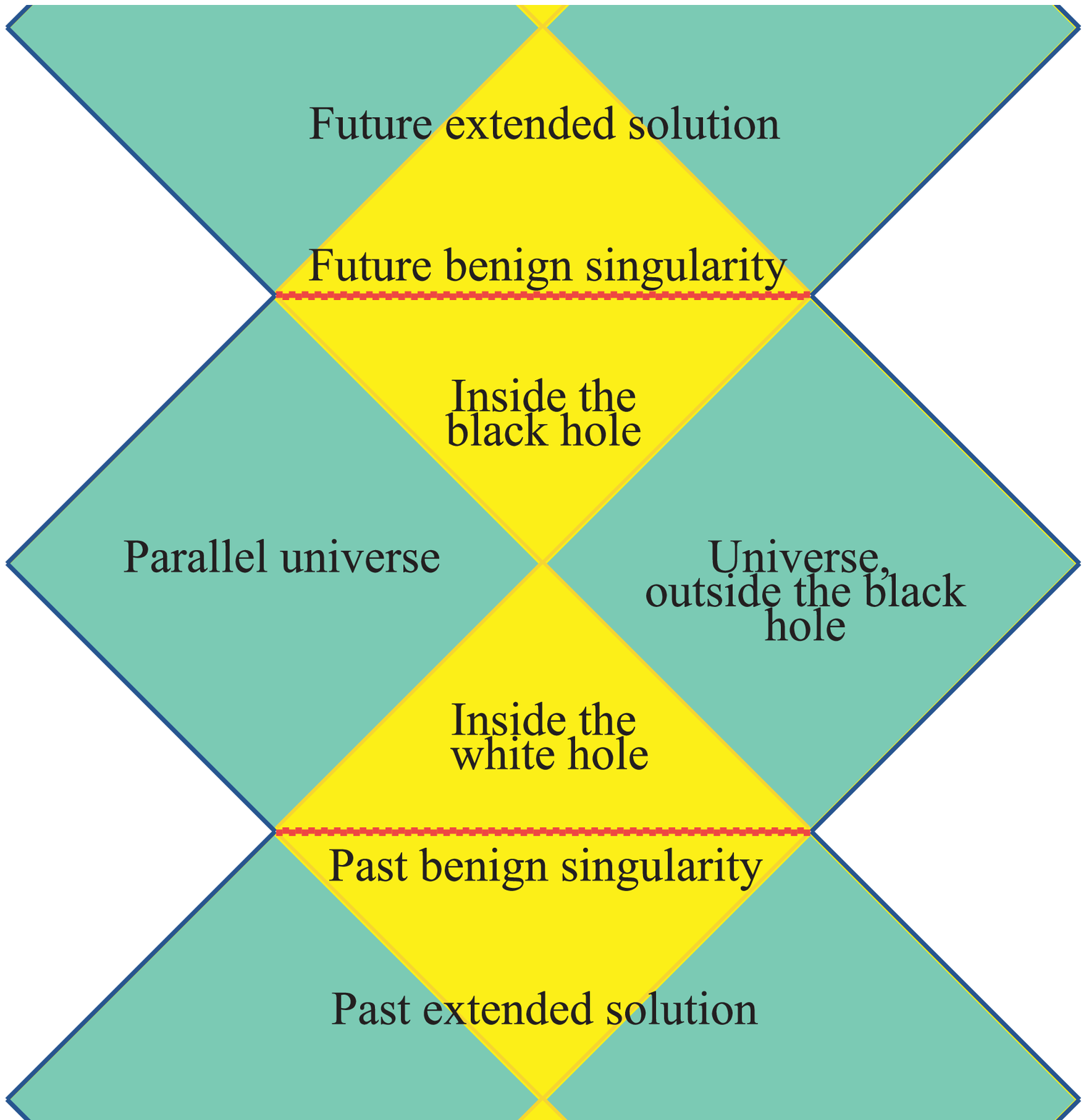}{0.5}{{\schw} solution, analytically extended beyond the $r=0$ singularity.}

This solution can be foliated in space+time, and therefore is globally hyperbolic.

\subsection{Space-like foliation of the {\rn} solution}

Fig. \ref{std-rn} shows the standard Penrose diagrams for the {\rn} spacetimes \cite{HE95}.

\image{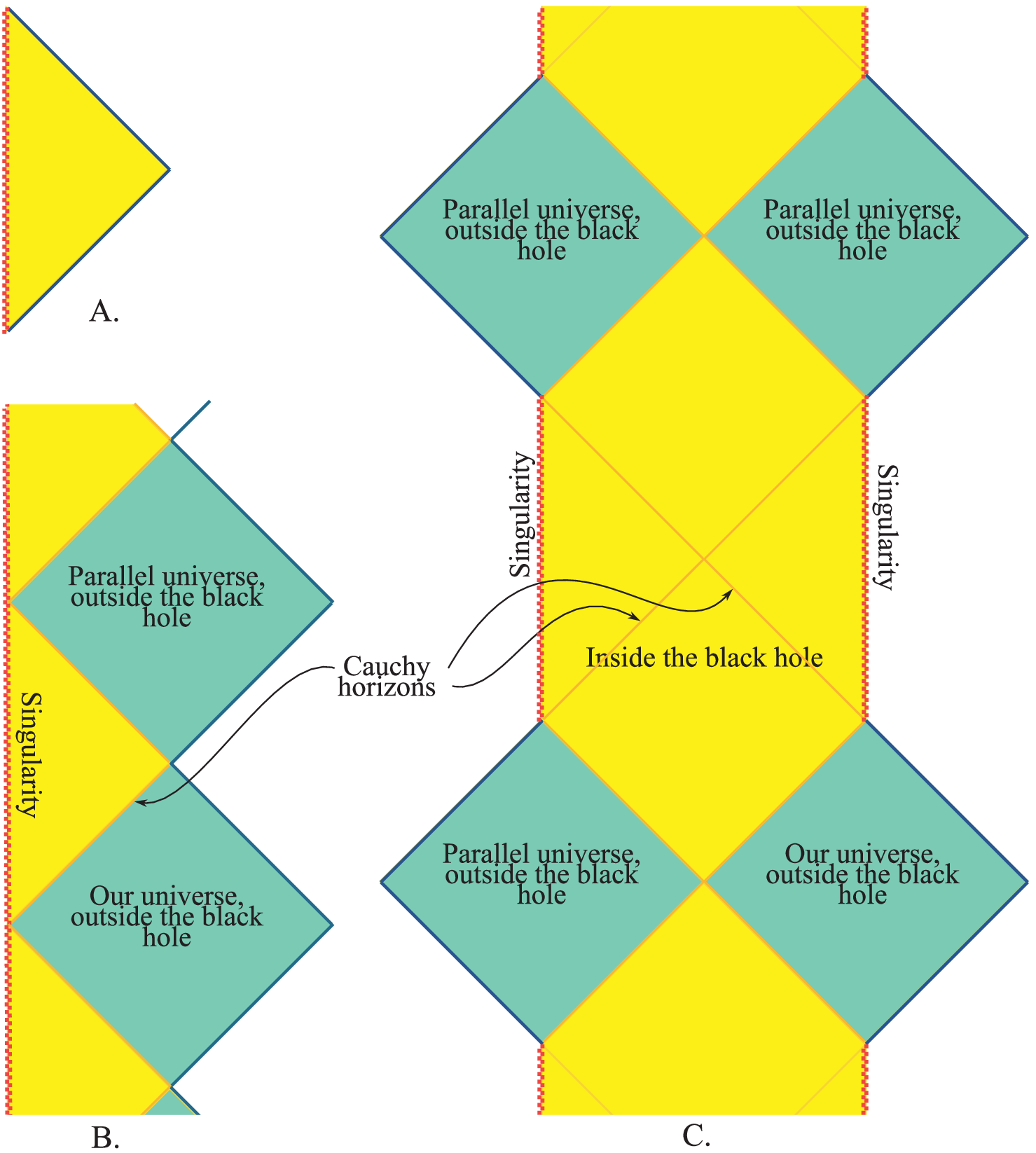}{0.7}{Reissner-Nordstr\"om black holes. A. Naked solutions ($q^2>m^2$). B. Extremal solution ($q^2=m^2$). C. Solutions with $q^2<m^2$.}

The Penrose diagrams \ref{rn-ext} shows how our extensions beyond the singularities allows the {\rn} solutions to be foliated in Cauchy hypersurfaces. In fig. \ref{rn-ext} B and C, in addition to extending the solution beyond the singularity, we cut out the spacetime along the Cauchy horizons. This is justified if the black holes form by collapse at a finite time, and then evaporate after a finite lifetime \cite{Sto11f,Sto12e}.

\image{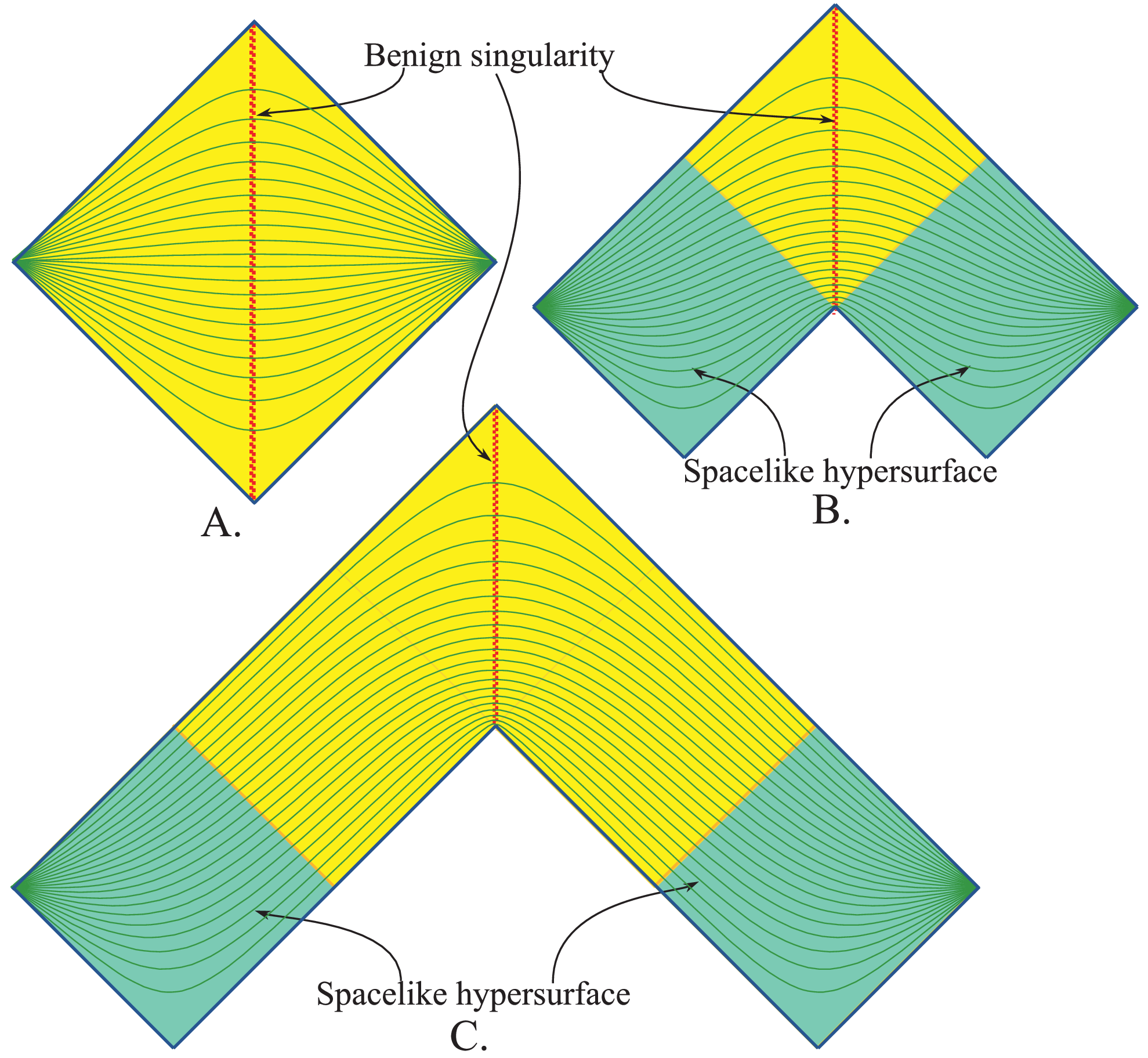}{0.75}{Reissner-Nordstr\"om black hole solutions, extended beyond the singularities, and restricted to globally hyperbolic regions. A. Naked solutions ($q^2>m^2$). B. Extremal solution ($q^2=m^2$). C. Solutions with $q^2<m^2$.}

For the {\kn} black holes, the foliations are similar to those for the {\rn} solutions \cite{Sto12e}, especially because the extension proposed in \cite{Sto11g} can be chosen so that the closed timelike curves disappear.

\subsection{Black hole information paradox}
\label{s_bh_info}

Bekenstein and Hawking discovered that black holes obey laws similar to those of thermodynamics, and proposed that these laws are in fact thermodynamics (see \cite{bekenstein1973black,bardeen1973four,HP96}, also \cite{strominger95houches,jacobson1996introductory} and references therein). Hawking realized that black holes evaporate, and the radiation is thermal. This led him to the idea that, after evaporation, the information is lost \cite{hawking1974bhexplosions,Haw75,Haw76}. Many solutions were proposed, such as \cite{susskind1993stretched}, \cite{Haw05}, \cite{preskill1993bh-info,page1994bh-info,banks1995bh-info,singh2004quantum-bh,prester2013curing}, \cite{corda2012effective,corda2011effective,corda2013black,corda2013effective}, \cite{almheiri2013black,hwang2012firewall,marolf2013gauge}, {\etc} It was proposed that quantum gravity would naturally cure this problem, but it has been suggested that in fact it would make the problem exist even in the absence of black holes \cite{itzhaki1995information}.

Since the extended {\schw} solution can be foliated in space+time (sections \sref{s_schw} and \sref{s_schw_foliation}), it can be used to represent evaporating electrically neutral non-rotating black holes. The solution can be analytically extended beyond $r=0$, hence the affirmation that the information is lost at the singularity is no longer supported. In fig. \ref{evaporating-bh-s} can be seen that our solution extends through the singularity, and allows the existence of globally hyperbolic spacetimes containing evaporating black holes.

\image{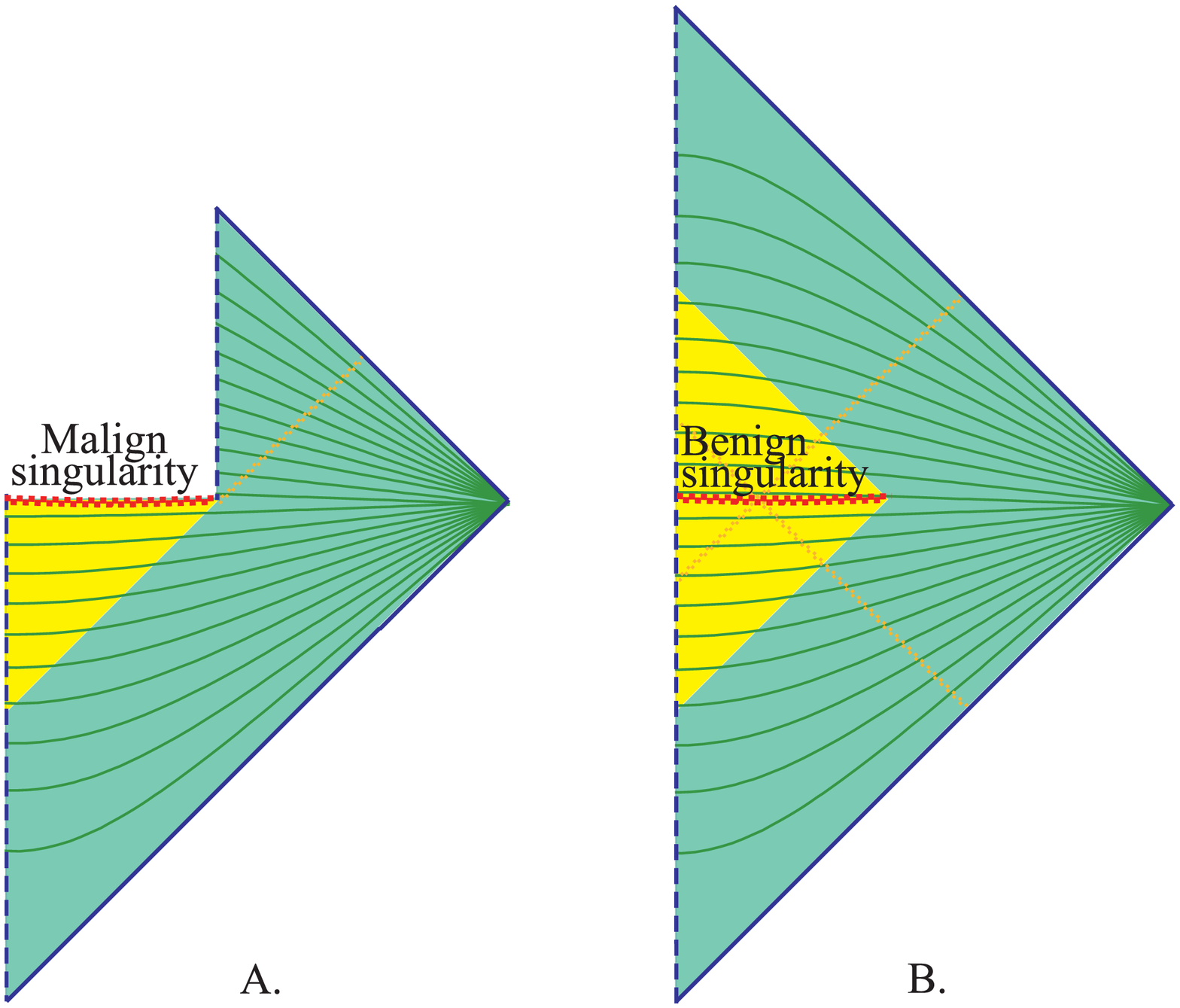}{0.6}{\textbf{A.} Standard evaporating black hole, whose singularity destroys the information.
\textbf{B.} Evaporating black hole extended through the singularity preserves information and admits a space+time foliation.}

\section{Possible experimental consequences and Quantum Gravity}

\subsection{Can we do experiments with singularities?}

We reviewed the foundations of Singular General Relativity (SGR), and its applications to black hole singularities.
SGR is a natural extension of GR, but nevertheless, it would be great to be able to submit it to experimental tests.
We have seen that the solutions are the same as those predicted by Einstein's equation, as long as the metric is {\nondeg}. The only differences appear where the metric is degenerate, at singularities. But how can we go to the singularities, or how can we generate singularities, and test the results at the singularities? How could we design an experimental apparatus which is not destroyed by the singularity? It seems that a direct experiment to test the predictions of SGR is not possible.

What about indirect tests? For example, if information is preserved, this would be an evidence in favor of SGR. But how can we test this? Can we monitor a black hole, from the time when it is formed, to the time when it evaporates completely, and check that the information is preserved during this entire process? The current knowledge predicts that this information will be anyway extremely scrambled. Even if we would be able to do this someday, the conservation of information is predicted by a long list of other approaches to Hawking's information loss paradox (see section \sref{s_bh_info}).

In General Relativity, classical elementary particles can be considered small black holes. If they are point-like, and have definite trajectories, then they are singularities, like the {\schw}, {\rn}, and {\kn} singularities. To go from classical to quantum, one applies path integrals over the classical trajectories. In this way, possible effects of the singularities may also be present at the points where the metric is non-singular.

In \cite{Sto12d} we suggested that the geometric and topological properties we identified at singularities have implications to Quantum Gravity (QG), as we shall see in the following. This suggests that it might be possible to test our approach by QG effects. One feature that seems to be required by most, if not all approaches to QG, is dimensional reduction. Singular General Relativity shows that singularities are accompanied in a natural way by dimensional reduction.

\subsection{Dimensional reduction in QFT and QG}

Various results obtained in Quantum Field Theory (QFT) and in QG suggest that at small scales a dimensional reduction should take place. The definition and the cause of this reduction differs from one approach to another. Here is just a small part of the literature using one form of dimensional reduction or another to obtain regularization in QFT and QG:
\begin{itemize}
	\item 
	Fractal universe \cite{Calc2010FractalQFT,Calc2010FractalUniverse}, based on a Lebesgue-Stieltjes measure or a fractional measure \cite{Calc2011FractalGeometry}, fractional calculus, and fractional action principles \cite{el2005fractional,el2007geom,el2007schw,el2007cosmology,el2008fractional,udriste2008euler,el2010modifications,el2012fractional,el2012gravitons,el2013fractional}.
	\item 
	Topological dimensional reduction \cite{shirkov2010coupling,FS2011KG,Fiz2010Riem,FS2012Axial,shirkov2012dreamland}.
	\item 
	Vanishing Dimensions at LHC \cite{ADFLS10}.
	\item 
	Dimensional reduction in QG \cite{Car95,Car09SDR,Car10sssst}.
	\item 
	Asymptotic safety \cite{Wein79AS}.
	\item 
	{\HL} gravity \cite{Hor09qglp}.
	\item 
	Other approaches to Quantum Gravity based on dimensional reduction include \cite{oda1997quantum,umetsu2010tunneling,moffat2010lorentz,mureika2012self,mureika2012primordial,charmousis2012higher}.\end{itemize}

Some of these types of dimensional reduction are very similar to those predicted by SGR to occur at benign singularities.

\subsection{Is dimensional reduction due to the benign singularities?}

Quantum Gravity is perturbatively non-renormalizable, but it can be made renormalizable by assuming one kind or another of dimensional reduction. The above mentioned approaches did this, by modifying General Relativity. In this section we point that several types of dimensional reduction which were postulated by various authors, occur naturally at our {\semireg} and {\quasireg} singularities \cite{Sto12d}.

\subsubsection{Geometric dimensional reduction}

	First, at each point where the metric becomes degenerate, a geometric, or {\em metric reduction} takes place, because the rank of the metric is reduced:
	\begin{equation}
	\dim{\coannih{T_p}M}=\dim{\annih{T_p}M}=\rank g_p.
\end{equation}

\subsubsection{Topological dimensional reduction}

From the Kupeli theorem \cite{Kup87b} follows that for constant signature, the manifold is locally a {\em product} $M=P\times_0 N$ between a manifold of lower dimension $P$ and another manifold $N$ with metric $0$. In other words, from the viewpoint of geometry, a region where the metric is degenerate and has constant signature can be identified with a lower dimensional space. This suggests a connection with the {\em topological dimensional reduction} explored by D.V. Shirkov and P. Fiziev \cite{shirkov2010coupling,FS2011KG,Fiz2010Riem,FS2012Axial,shirkov2012dreamland}.

\subsubsection{Vanishing of gravitons}

	If the singularity is {\quasireg}, the Weyl tensor $C_{abcd} \to 0$ as approaching a {\quasireg} singularity. This implies that the {\em local degrees of freedom} -- \ie the gravitational waves for GR and the gravitons for QG -- vanish, allowing by this the needed renormalizability \cite{Car95}.

\subsubsection{Anisotropy between space and time}

In \cite{Sto11f} we obtained new coordinates, which make the {\rn} metric analytic at the singularity. In these coordinates, the metric is given by equation \eqref{eq_rn_ext_ext}.
A {\em charged particle} with spin $0$ can be viewed, at least classically, as a {\rn} black hole. The above metric reduces its dimension to dim $=2$.

To admit space+time foliation in these coordinates, we should take {\em $T\geq 3S$}. An open research problem is whether this anisotropy is connected to the similar anisotropy from {\em \HL} gravity, introduced in \cite{Hor09qglp}.

\subsubsection{Measure dimensional reduction}

In the {\em fractal universe approach} \cite{Calc2010FractalQFT,Calc2010FractalUniverse,Calc2011FractalGravity}, one expresses the measure in the integral
\begin{equation}
S=\int_\mc M\de\varrho(x)\, \mc L,
\end{equation}
in terms of some functions $f_{(\mu)}(x)$, some of them vanishing at low scales:
\begin{equation}
\label{eq_stme}
\de \varrho(x) = \prod_{\mu=0}^{D-1} f_{(\mu)}(x)\,\de x^\mu.
\end{equation}

In {\em Singular General Relativity},
\begin{equation}
\label{eq_stme_sgr}
\de \varrho(x) = \sqrt{-\det g}\de x^D.
\end{equation}
If the metric is diagonal in the coordinates $(x^\mu)$, then we can take
\begin{equation}
\label{eq_stme_sgr_weights}
f_{(\mu)}(x) = \sqrt{\abs{g_{\mu\mu}(x)}}.
\end{equation}

This suggests that the results obtained by Calcagni by considering the universe to be fractal follow naturally from the benign metrics.

\subsection{Dimensional reduction and Quantum Gravity}

The Singular General Relativity approach leads, as a side effect, to various types of dimensional reduction, which are similar to those proposed in the literature to make Quantum Gravity perturbatively renormalizable. By investigating the non-renormalizability problems appearing when quantizing gravity, many researchers were led to the conclusion that the problem would vanish if one kind of dimensional reduction or another is postulated (sometimes ad-hoc). By contrary, our approach led to this as a natural consequence of understanding the singularities.

Of course, in SGR the dimensional reduction appears at the singularity, while QG is expected to be perturbatively renormalizable everywhere. But if classical particles are singularities, quantum particles behave like sums over histories of classical particles. Thus, at any point there will be virtual singularities to contribute to the Feynman integrals. 
This means that the effects will be present everywhere. They are expected as a reduction of the determinant of the metric, and of the Weyl curvature tensor, which allows the desired regularization. Moreover, as the energy increases, the order of the Feynman diagrams in the same region increases, and we expect that the dimensional reduction effects induced by singularities becomes more significant too. It is an open question at this time whether this dimensional reduction is enough to regularize gravity, but this research is just at the beginning.

\section{Conclusions}

We reviewed some of our results of Singular General Relativity \cite{Sto13a}, concerning the black hole singularities. Some singularities allow the canonical and invariant construction of geometric objects which remain smooth and non-singular. By using these objects, one can write equations which are equivalent to Einstein's equations outside singularities, but in addition extend smoothly at singularities. The {\FLRW} big bang singularities turn out to be of this type. The black hole singularities can be made so by removing the coordinate singularity. For the charged black hole singularities, the electromagnetic potential and field become smooth. The singularities of the black hole having a finite life span are compatible with global hyperbolicity and conservation of information. Such singularities are accompanied by dimensional reduction, a feature which is desired by many approaches to Quantum Gravity. While in these approaches dimensional reduction is obtained by modifying General Relativity, these singularities lead naturally to it, within the framework of GR.

There is a rich literature concerning gravity, black holes and singularities in lower or higher dimensions (see {\eg} \cite{brown1988lower,strominger95houches,emparan2008bh-high,Watcharangkool2012} and references therein). 
While the geometric apparatus of Singular {\ssemiriem} Geometry reviewed in section 
\sref{s_ssrg} works for other dimensions too, in this review we focused only on four-dimensional spacetimes, and some of the results don't work in more dimensions.

\subsection*{Acknowledgements}
The author thanks an anonymous referee for the valuable suggestions to improve the completeness of this review.


\begin{thebibliography}{CKGDS09}

\bibitem[ACBY00]{ACY00}
A.~Anderson, Y.~Choquet-Bruhat, and J.~York, \emph{{Einstein's Equations and
  Equivalent Hyperbolic Dynamical Systems}}, Mathematical and Quantum Aspects
  of Relativity and Cosmology (2000), 30--54.

\bibitem[ADF{\etalchar{+}}12]{ADFLS10}
L.~Anchordoqui, D.~C. Dai, M.~Fairbairn, G.~Landsberg, and D.~Stojkovic,
  \emph{Vanishing dimensions and planar events at the {LHC}}, Mod. Phys. Lett.
  A \textbf{27} (2012), no.~04.

\bibitem[ADM62]{ADM62}
R.~Arnowitt, S.~Deser, and C.~W. Misner, \emph{{The Dynamics of General
  Relativity}}, {Gravitation: An Introduction to Current Research}, Wiley, New
  York, 1962, pp.~227--264.

\bibitem[AMPS13]{almheiri2013black}
A.~Almheiri, D.~Marolf, J.~Polchinski, and J.~Sully, \emph{Black holes:
  complementarity or firewalls?}, Journal of High Energy Physics \textbf{2013}
  (2013), no.~2, 1--20.

\bibitem[AS11]{ashtekar2011LQC}
A.~Ashtekar and P.~Singh, \emph{Loop quantum cosmology: a status report},
  Class. Quant. Grav. \textbf{28} (2011), no.~21, 213001--213122,
  \href{http://arxiv.org/abs/1108.0893}{arXiv:gr-qc/1108.0893}.

\bibitem[AT99a]{AT99i}
K.~Anguige and K.~P. Tod, \emph{Isotropic cosmological singularities: {I}.
  {P}olytropic perfect fluid spacetimes}, Ann. of Phys. \textbf{276} (1999),
  no.~2, 257--293.

\bibitem[AT99b]{AT99ii}
\bysame, \emph{Isotropic cosmological singularities: {II}. {T}he
  {E}instein-{V}lasov system}, Ann. of Phys. \textbf{276} (1999), no.~2,
  294--320.

\bibitem[Ban95]{banks1995bh-info}
Tom Banks, \emph{Lectures on black holes and information loss}, Nuclear Physics
  B-Proceedings Supplements \textbf{41} (1995), no.~1, 21--65.

\bibitem[BCH73]{bardeen1973four}
J.~M. Bardeen, B.~Carter, and S.~W. Hawking, \emph{The four laws of black hole
  mechanics}, Comm. Math. Phys. \textbf{31} (1973), no.~2, 161--170.

\bibitem[Bek73]{bekenstein1973black}
J.~D. Bekenstein, \emph{Black holes and entropy}, Phys. Rev. D \textbf{7}
  (1973), no.~8, 2333.

\bibitem[Bes87]{BESS87}
Arthur~L. Besse, \emph{{E}instein manifolds, {E}rgebnisse der {M}athematik und
  ihrer {G}renzgebiete (3) [{R}esults in mathematics and related areas (3)],
  vol. 10}, Berlin, New York: Springer-Verlag, 1987.

\bibitem[BJZ12]{borisov2012fR}
Alexander Borisov, Bhuvnesh Jain, and Pengjie Zhang, \emph{Spherical collapse
  in {f(R)} gravity}, Phys. Rev. D \textbf{85} (2012), no.~6, 063518.

\bibitem[Boj01]{bojowald2003absenceLQC}
M.~Bojowald, \emph{{Absence of a Singularity in Loop Quantum Cosmology}}, Phys.
  Rev. Lett. \textbf{86} (2001), no.~23, 5227--5230,
  \href{http://arxiv.org/abs/gr-qc/0102069}{arXiv:gr-qc/0102069}.

\bibitem[Bro88]{brown1988lower}
J~David Brown, \emph{Lower dimensional gravity}, World Scientific, 1988.

\bibitem[BS97]{barve1997naked}
Sukratu Barve and TP~Singh, \emph{Are naked singularities forbidden by the
  second law of thermodynamics?}, Modern Physics Letters A \textbf{12} (1997),
  no.~32, 2415--2419.

\bibitem[BSVW98a]{barve1998particle}
Sukratu Barve, TP~Singh, Cenalo Vaz, and Louis Witten, \emph{Particle creation
  in the marginally bound, self-similar collapse of inhomogeneous dust},
  Nuclear physics B \textbf{532} (1998), no.~1, 361--375.

\bibitem[BSVW98b]{barve1998quantum}
\bysame, \emph{Quantum stress tensor in self-similar spherical dust collapse},
  Physical Review D \textbf{58} (1998), no.~10, 104018.

\bibitem[BSW00]{barve2000spherical}
Sukratu Barve, TP~Singh, and Louis Witten, \emph{Spherical gravitational
  collapse: Tangential pressure and related equations of state}, General
  Relativity and Gravitation \textbf{32} (2000), no.~4, 697--717.

\bibitem[Buc70]{buchdahl1970fR}
Hans~A Buchdahl, \emph{Non-linear lagrangians and cosmological theory}, Monthly
  Notices of the Royal Astronomical Society \textbf{150} (1970), 1.

\bibitem[Cal10a]{Calc2010FractalUniverse}
G.~Calcagni, \emph{{Fractal universe and quantum gravity}}, Phys. Rev. Lett.
  \textbf{104} (2010), no.~25, 251301,
  \href{http://arxiv.org/abs/0912.3142}{arXiv:hep-th/0912.3142}.

\bibitem[Cal10b]{Calc2010FractalQFT}
\bysame, \emph{{Quantum field theory, gravity and cosmology in a fractal
  universe}}, Journal of High Energy Physics \textbf{2010} (2010), no.~3,
  1--38, \href{http://arxiv.org/abs/1001.0571}{arXiv:hep-th/1001.0571}.

\bibitem[Cal11a]{Calc2011FractalGeometry}
\bysame, \emph{Geometry of fractional spaces},
  \href{http://arxiv.org/abs/1106.5787}{arXiv:hep-th/1106.5787} (2011).

\bibitem[Cal11b]{Calc2011FractalGravity}
\bysame, \emph{Gravity on a multifractal}, Physics Letters B (2011),
  \href{http://arxiv.org/abs/1012.1244}{arXiv:hep-th/1012.1244}.

\bibitem[Car95]{Car95}
S.~Carlip, \emph{{Lectures in (2+ 1)-dimensional gravity}}, J. Korean Phys. Soc
  \textbf{28} (1995), S447--S467,
  \href{http://arxiv.org/abs/gr-qc/9503024}{arXiv:gr-qc/9503024}.

\bibitem[Car10]{Car10sssst}
\bysame, \emph{{The Small Scale Structure of Spacetime}},
  \href{http://arxiv.org/abs/1009.1136}{arXiv:gr-qc/1009.1136} (2010).

\bibitem[CBY02]{CnY02}
Y.~Choquet-Bruhat and J.~York, \emph{{Constraints and evolution in cosmology}},
  Cosmological crossroads (2002), 29--58.

\bibitem[CC10]{corda2010removingBHsingularities}
C.~Corda and H.~J.~M. Cuesta, \emph{{Removing Black Hole Singularities with
  Nonlinear Electrodynamics}}, Mod. Phys. Lett. A \textbf{25} (2010), no.~28,
  2423--2429, \href{http://arxiv.org/abs/0905.3298v8}{arXiv:gr-qc/0905.3298}.

\bibitem[CGK12]{charmousis2012higher}
C~Charmousis, B~Gout{\'e}raux, and E~Kiritsis, \emph{Higher-derivative
  scalar-vector-tensor theories: black holes, galileons, singularity cloaking
  and holography}, Journal of High Energy Physics \textbf{2012} (2012), no.~9,
  1--44.

\bibitem[CHKS13]{corda2013effective}
C.~Corda, S.H/ Hendi, R.~Katebi, and N.O. Schmidt, \emph{Effective state,
  {H}awking radiation and quasi-normal modes for {K}err black holes}, Journal
  of High Energy Physics \textbf{2013} (2013), no.~6, 1--12.

\bibitem[CKGDS09]{Car09SDR}
S.~Carlip, J.~Kowalski-Glikman, R.~Durka, and M.~Szczachor, \emph{Spontaneous
  dimensional reduction in short-distance quantum gravity?}, AIP Conference
  Proceedings, vol.~31, 2009, p.~72.

\bibitem[CN98]{CN98}
C.~M. Claudel and K.~P. Newman, \emph{{The Cauchy problem for quasi--linear
  hyperbolic evolution problems with a singularity in the time}}, P. Roy. Soc.
  A-Math. Phy. \textbf{454} (1998), no.~1972, 1073.

\bibitem[Cor11]{corda2011effective}
C.~Corda, \emph{Effective temperature for black holes}, Journal of High Energy
  Physics \textbf{2011} (2011), no.~8, 1--10.

\bibitem[Cor12]{corda2012effective}
\bysame, \emph{Effective temperature, {H}awking radiation and quasinormal
  modes}, International Journal of Modern Physics D \textbf{21} (2012), no.~11.

\bibitem[Cor13]{corda2013black}
\bysame, \emph{Black hole quantum spectrum}, The European Physical Journal C
  \textbf{73} (2013), no.~12, 1--12.

\bibitem[Edd24]{eddington1924comparison}
A.~S. Eddington, \emph{{A Comparison of Whitehead's and Einstein's Formulae}},
  Nature \textbf{113} (1924), 192.

\bibitem[EN05]{el2005fractional}
R.A. El-Nabulsi, \emph{A fractional action-like variational approach of some
  classical, quantum and geometrical dynamics}, International Journal of
  Applied Mathematics \textbf{17} (2005), no.~3, 299.

\bibitem[EN07a]{el2007cosmology}
\bysame, \emph{Cosmology with a fractional action principle}, Rom. Rep. Phys.
  \textbf{59} (2007), no.~3, 759--765.

\bibitem[EN07b]{el2007geom}
\bysame, \emph{Differential geometry and modern cosmology with fractionaly
  differentiated lagrangian function and fractional decaying force term}, Rom.
  Journ. Phys. \textbf{52} (2007), no.~3--4, 467.

\bibitem[EN07c]{el2007schw}
\bysame, \emph{Some fractional geometrical aspects of weak field approximation
  and schwarzschild spacetime}, Rom. Journ. Phys. \textbf{52} (2007), no.~5--7,
  705--715.

\bibitem[EN10]{el2010modifications}
\bysame, \emph{Modifications at large distances from fractional and fractal
  arguments}, Fractals \textbf{18} (2010), no.~02, 185--190.

\bibitem[EN12]{el2012gravitons}
\bysame, \emph{Gravitons in fractional action cosmology}, International Journal
  of Theoretical Physics \textbf{51} (2012), no.~12, 3978--3992.

\bibitem[EN13]{el2013fractional}
\bysame, \emph{Fractional derivatives generalization of einstein's field
  equations}, Indian Journal of Physics \textbf{87} (2013), no.~2, 195--200.

\bibitem[ENT08]{el2008fractional}
R.A. El-Nabulsi and D.F.M. Torres, \emph{Fractional actionlike variational
  problems}, Journal of Mathematical Physics \textbf{49} (2008), no.~5,
  053521--053521.

\bibitem[ENW12]{el2012fractional}
R.A. El-Nabulsi and C.G. Wu, \emph{Fractional complexified field theory from
  {S}axena-{K}umbhat fractional integral, fractional derivative of order
  $(\alpha,\beta)$ and dynamical fractional integral exponent}, African
  Diaspora Journal of Mathematics. New Series \textbf{13} (2012), no.~1,
  45--61.

\bibitem[ER08]{emparan2008bh-high}
Roberto Emparan and Harvey~S Reall, \emph{Black holes in higher dimensions},
  Living Rev. Rel. \textbf{11} (2008), no.~6, 0801--3471.

\bibitem[FB52]{ChoquetBruhat1952CauchyGR}
Y.~Foures-Bruhat, \emph{Th{\'e}or{\`e}me d'existence pour certains syst{\`e}mes
  d'{\'e}quations aux d{\'e}riv{\'e}es partielles non lin{\'e}aires}, Acta
  Mathematica \textbf{88} (1952), no.~1, 141--225.

\bibitem[Fin58]{finkelstein1958past}
D.~Finkelstein, \emph{Past-future asymmetry of the gravitational field of a
  point particle}, Phys. Rev. \textbf{110} (1958), no.~4, 965.

\bibitem[Fiz10]{Fiz2010Riem}
P.~P. Fiziev, \emph{{{R}iemannian (1+d)-Dim Space-Time Manifolds with
  Nonstandard Topology which Admit Dimensional Reduction to Any Lower Dimension
  and Transformation of the Klein-Gordon Equation to the 1-Dim Schr\"odinger
  Like Equation}},
  \href{http://arxiv.org/abs/1012.3520}{arXiv:math-ph/1012.3520} (2010).

\bibitem[Fri22]{FRI22de}
A.~Friedman, \emph{{{\"U}ber die Kr{\"u}mmung des Raumes}}, Zeitschrift f{\"u}r
  Physik A Hadrons and Nuclei \textbf{10} (1922), no.~1, 377--386.

\bibitem[Fri24]{FRI24}
\bysame, \emph{{{\"U}ber die M{\"o}glichkeit einer Welt mit konstanter
  negativer Kr{\"u}mmung des Raumes}}, Zeitschrift f{\"u}r Physik A Hadrons and
  Nuclei \textbf{21} (1924), no.~1, 326--332.

\bibitem[Fri99]{FRI99en}
\bysame, \emph{{On the Curvature of Space}}, General Relativity and Gravitation
  \textbf{31} (1999), no.~12, 1991--2000.

\bibitem[FS11]{FS2011KG}
P.~P. Fiziev and D.~V. Shirkov, \emph{Solutions of the {K}lein-{G}ordon
  equation on manifolds with variable geometry including dimensional
  reduction}, Theoretical and Mathematical Physics \textbf{167} (2011), no.~2,
  680--691, \href{http://arxiv.org/abs/1009.5309}{arXiv:hep-th/1009.5309}.

\bibitem[FS12]{FS2012Axial}
\bysame, \emph{{The (2+1)-dim Axial Universes -- Solutions to the Einstein
  Equations, Dimensional Reduction Points, and Klein--Fock--Gordon Waves}}, J.
  Phys. A \textbf{45} (2012), no.~055205, 1--15,
  \href{http://arxiv.org/abs/1104.0903}{arXiv:gr-qc/arXiv:1104.0903}.

\bibitem[GHL04]{GHLF04}
S.~Gallot, D.~Hullin, and J.~Lafontaine, \emph{{R}iemannian geometry}, 3rd ed.,
  Springer-Verlag, Berlin, New York, 2004.

\bibitem[GS86]{GS86uvgr}
M.~H. Goroff and A.~Sagnotti, \emph{{The ultraviolet behavior of Einstein
  gravity}}, Nuclear Physics B \textbf{266} (1986), no.~3-4, 709--736.

\bibitem[Haw66a]{Haw66i}
S.~W. Hawking, \emph{The occurrence of singularities in cosmology}, P. Roy.
  Soc. A-Math. Phy. \textbf{294} (1966), no.~1439, 511--521.

\bibitem[Haw66b]{Haw66ii}
\bysame, \emph{The occurrence of singularities in cosmology. {II}}, P. Roy.
  Soc. A-Math. Phy. \textbf{295} (1966), no.~1443, 490--493.

\bibitem[Haw67]{Haw67iii}
\bysame, \emph{The occurrence of singularities in cosmology. {III}. {C}ausality
  and singularities}, P. Roy. Soc. A-Math. Phy. \textbf{300} (1967), no.~1461,
  187--201.

\bibitem[Haw74]{hawking1974bhexplosions}
Stephen~W Hawking, \emph{Black hole explosions}, Nature \textbf{248} (1974),
  no.~5443, 30--31.

\bibitem[Haw75]{Haw75}
S.~W. Hawking, \emph{{Particle Creation by Black Holes}}, Comm. Math. Phys.
  \textbf{43} (1975), no.~3, 199--220.

\bibitem[Haw76]{Haw76}
\bysame, \emph{{Breakdown of Predictability in Gravitational Collapse}}, Phys.
  Rev. D \textbf{14} (1976), no.~10, 2460.

\bibitem[Haw05]{Haw05}
\bysame, \emph{{Information Loss in Black Holes}}, Phys. Rev. D \textbf{72}
  (2005), no.~8, 084013,
  \href{http://arxiv.org/abs/hep-th/0507171}{arXiv:hep-th/0507171}.

\bibitem[HE95]{HE95}
S.~W. Hawking and G.~F.~R. Ellis, \emph{{The Large Scale Structure of Space
  Time}}, Cambridge University Press, 1995.

\bibitem[HLY12]{hwang2012firewall}
Dong-il Hwang, Bum-Hoon Lee, and Dong-han Yeom, \emph{Is the firewall
  consistent?}, arXiv preprint arXiv:1210.6733 (2012),
  \href{http://arxiv.org/abs/1210.6733}{arXiv:1210.6733}.

\bibitem[HM11]{hendi2011black}
S.H. Hendi and D.~Momeni, \emph{Black-hole solutions in {f(R)} gravity with
  conformal anomaly}, Eur. Phys. J. C \textbf{71} (2011), no.~12, 1--9.

\bibitem[Ho{\v{r}}09]{Hor09qglp}
P.~Ho{\v{r}}ava, \emph{{Quantum Gravity at a Lifshitz Point}}, Phys. Rev. D
  \textbf{79} (2009), no.~8, 084008,
  \href{http://arxiv.org/abs/0901.3775}{arXiv:hep-th/0901.3775}.

\bibitem[HP70]{HP70}
S.~W. Hawking and R.~W. Penrose, \emph{{The Singularities of Gravitational
  Collapse and Cosmology}}, Proc. Roy. Soc. London Ser. A \textbf{314} (1970),
  no.~1519, 529--548.

\bibitem[HP96]{HP96}
\bysame, \emph{{The Nature of Space and Time}}, Princeton University Press,
  Princeton and Oxford, 1996.

\bibitem[IPS81]{isham1981qg2}
Chris~J. Isham, Roger Penrose, and Dennis~William Sciama, \emph{Quantum gravity
  2}, Quantum Gravity II, vol.~1, 1981.

\bibitem[Itz95]{itzhaki1995information}
N.~Itzhaki, \emph{Information loss in quantum gravity without black holes},
  Classical and Quantum Gravity \textbf{12} (1995), no.~11, 2747.

\bibitem[Jac96]{jacobson1996introductory}
Ted Jacobson, \emph{Introductory lectures on black hole thermodynamics},
  Lectures given at the University of Utrecht, The Netherlands (1996),
  \href{http://www.physics.umd.edu/grt/taj/776b/lectures.pdf}{http://www.physics.umd.edu/grt/taj/776b/lectures.pdf}.

\bibitem[Jos13]{joshi2013spacetime}
Pankaj~S Joshi, \emph{Spacetime singularities}, arXiv preprint arXiv:1311.0449
  (2013), \href{http://arxiv.org/abs/1311.0449}{arXiv:1311.0449}.

\bibitem[Kup87]{Kup87b}
D.~Kupeli, \emph{Degenerate manifolds}, Geom. Dedicata \textbf{23} (1987),
  no.~3, 259--290.

\bibitem[Lem27]{LEM27}
G.~Lema{\^\i}tre, \emph{Un univers homog{\`e}ne de masse constante et de rayon
  croissant rendant compte de la vitesse radiale des n{\'e}buleuses
  extra-galactiques}, Annales de la Societe Scietifique de Bruxelles
  \textbf{47} (1927), 49--59.

\bibitem[MN12]{mureika2012self}
J.~Mureika and P.~Nicolini, \emph{Self-completeness and spontaneous dimensional
  reduction}, arXiv preprint arXiv:1206.4696 (2012),
  \href{http://arxiv.org/abs/1206.4696}{arXiv:1206.4696}.

\bibitem[Mof10]{moffat2010lorentz}
JW~Moffat, \emph{Lorentz violation of quantum gravity}, Classical and Quantum
  Gravity \textbf{27} (2010), no.~13, 135016.

\bibitem[MP13]{marolf2013gauge}
D.~Marolf and J.~Polchinski, \emph{Gauge-gravity duality and the black hole
  interior}, Physical review letters \textbf{111} (2013), no.~17, 171301.

\bibitem[Mur12]{mureika2012primordial}
J.~Mureika, \emph{Primordial black hole evaporation and spontaneous dimensional
  reduction}, Physics letters. Section B \textbf{716} (2012), no.~1, 171--175.

\bibitem[Oda97]{oda1997quantum}
Ichiro Oda, \emph{Quantum instability of black hole singularity in three
  dimensions}, Arxiv preprint gr-qc/9703056 (1997),
  \href{http://arxiv.org/abs/gr-qc/9703056}{arXiv:gr-qc/9703056}.

\bibitem[O'N83]{ONe83}
B.~O'Neill, \emph{Semi-{R}iemannian geometry with applications to relativity},
  Pure Appl. Math., no. 103, Academic Press, New York-London, 1983.

\bibitem[ORG11]{olmo2011palatini-fR}
Gonzalo~J Olmo and D~Rubiera-Garcia, \emph{Palatini {f(R)} black holes in
  nonlinear electrodynamics}, Physical Review D \textbf{84} (2011), no.~12,
  124059.

\bibitem[Pag94]{page1994bh-info}
Don~N Page, \emph{Black hole information}, Proceedings of the 5th Canadian
  Conference on General Relativity and Relativistic Astrophysics, vol.~1, 1994,
  p.~1.

\bibitem[Pen65]{Pen65}
R.~Penrose, \emph{{Gravitational Collapse and Space-Time Singularities}}, Phys.
  Rev. Lett. \textbf{14} (1965), no.~3, 57--59.

\bibitem[Pen69]{Pen69}
\bysame, \emph{{Gravitational Collapse: the Role of General Relativity}},
  Revista del Nuovo Cimento; Numero speciale 1 (1969), 252--276.

\bibitem[Pen78]{Pen78}
R.~Penrose, \emph{{Singularities of spacetime}}, {Theoretical Principles in
  Astrophysics and Relativity} (N.R. Lebovitz, W.H. Reid, and P.O. Vandervoort,
  eds.), vol.~1, University of Chicago Press, 1978, pp.~217--243.

\bibitem[Pen79]{Pen79}
\bysame, \emph{{Singularities and time-asymmetry}}, {General relativity: an
  Einstein centenary survey}, vol.~1, 1979, pp.~581--638.

\bibitem[Pen98]{Pen98}
R.~Penrose, \emph{{The Question of Cosmic Censorship}}, {(ed. R. M. Wald) Black
  Holes and Relativistic Stars}, University of Chicago Press, Chicago,
  Illinois, 1998, pp.~233--248.

\bibitem[Pre93]{preskill1993bh-info}
J~Preskill, \emph{Do black holes destroy information?}, Black Holes, Membranes,
  Wormholes and Superstrings, vol.~1, World Scientific, River Edge, NJ., 1993,
  p.~22.

\bibitem[Pre13]{prester2013curing}
Predrag~Dominis Prester, \emph{Curing black hole singularities with local scale
  invariance}, arXiv preprint arXiv:1309.1188 (2013).

\bibitem[Rob35]{ROB35I}
H.~P. Robertson, \emph{{Kinematics and World-Structure}}, The Astrophysical
  Journal \textbf{82} (1935), 284.

\bibitem[Rob36a]{ROB35II}
\bysame, \emph{{Kinematics and World-Structure II.}}, The Astrophysical Journal
  \textbf{83} (1936), 187.

\bibitem[Rob36b]{ROB35III}
\bysame, \emph{{Kinematics and World-Structure III.}}, The Astrophysical
  Journal \textbf{83} (1936), 257.

\bibitem[Rod06]{rodnianski2006cauchy}
Igor Rodnianski, \emph{The cauchy problem in general relativity}, Proceedings
  oh the International Congress of Mathematicians: Madrid, August 22-30, 2006:
  invited lectures, 2006, pp.~421--442.

\bibitem[Sch16a]{Scw16b}
K.~Schwarzschild, \emph{{{\"U}ber das Gravitationsfeld eines Kugel aus
  inkompressibler Fl{\"u}ssigkeit nach der Einsteinschen Theorie}},
  Sitzungsber. Preuss. Akad. D. Wiss. (1916), 424--434,
  \href{http://arxiv.org/abs/physics/9912033}{arXiv:physics/9912033}.

\bibitem[Sch16b]{Scw16a}
\bysame, \emph{{{\"U}ber das Gravitationsfeld eines Massenpunktes nach der
  Einsteinschen Theorie}}, Sitzungsber. Preuss. Akad. D. Wiss. (1916),
  189--196, \href{http://arxiv.org/abs/physics/9905030}{arXiv:physics/9905030}.

\bibitem[Sen98]{senovilla1998singularity}
Jos{\'e}~MM Senovilla, \emph{Singularity theorems and their consequences},
  General Relativity and Gravitation \textbf{30} (1998), no.~5, 701--848.

\bibitem[Shi10]{shirkov2010coupling}
D.~V. Shirkov, \emph{Coupling running through the looking-glass of dimensional
  reduction}, Phys. Part. Nucl. Lett. \textbf{7} (2010), no.~6, 379--383,
  \href{http://arxiv.org/abs/1004.1510}{arXiv:hep-th/1004.1510}.

\bibitem[Shi11]{shirkov2012dreamland}
\bysame, \emph{{Dream-land with Classic Higgs field, Dimensional Reduction and
  all that}}, {Proceedings of the Steklov Institute of Mathematics}, vol. 272,
  2011, pp.~216--222.

\bibitem[ST69]{ST69}
I.~M. Singer and J.~A. Thorpe, \emph{{The curvature of 4-dimensional Einstein
  spaces}}, {G}lobal Analysis ({P}apers in Honor of {K}. {K}odaira), {Princeton
  Univ. Press, Princeton, and Univ. Tokyo Press, Tokyo}, 1969, pp.~355--365.

\bibitem[Sta80]{starobinsky1980fR}
Alexei~A Starobinsky, \emph{A new type of isotropic cosmological models without
  singularity}, Phys. Rev. B \textbf{91} (1980), no.~1, 99--102.

\bibitem[Sto11a]{Sto11a}
O.~C. Stoica, \emph{On singular semi-{R}iemannian manifolds}, To appear in Int.
  J. Geom. Methods Mod. Phys. (2011),
  \href{http://arxiv.org/abs/1105.0201}{arXiv:math.DG/1105.0201}.

\bibitem[Sto11b]{Sto11b}
\bysame, \emph{Warped products of singular semi-{R}iemannian manifolds}, Arxiv
  preprint math.DG/1105.3404 (2011),
  \href{http://arxiv.org/abs/1105.3404}{arXiv:math.DG/1105.3404}.

\bibitem[Sto12a]{Sto11f}
\bysame, \emph{Analytic {R}eissner-{N}ordstr{\"o}m singularity},
  \href{http://stacks.iop.org/1402-4896/85/i=5/a=055004}{Phys. Scr.}
  \textbf{85} (2012), no.~5, 055004,
  \href{http://arxiv.org/abs/1111.4332}{arXiv:gr-qc/1111.4332}.

\bibitem[Sto12b]{Sto12a}
\bysame, \emph{Beyond the {F}riedmann-{L}ema{\^i}tre-{R}obertson-{W}alker {B}ig
  {B}ang singularity}, Commun. Theor. Phys. \textbf{58} (2012), no.~4,
  613--616, \href{http://arxiv.org/abs/1203.1819}{arXiv:gr-qc/1203.1819}.

\bibitem[Sto12c]{Sto12e}
\bysame,
  \emph{\href{http://www.degruyter.com/view/j/auom.2012.20.issue-2/v10309-012-0050-3/v10309-012-0050-3.xml}{Spacetimes
  with Singularities}}, An. {\c S}t. Univ. Ovidius Constan{\c t}a \textbf{20}
  (2012), no.~2, 213--238,
  \href{http://arxiv.org/abs/1108.5099}{arXiv:gr-qc/1108.5099}.

\bibitem[Sto12d]{Sto12d}
\bysame, \emph{Quantum gravity from metric dimensional reduction at
  singularities}, Arxiv preprint gr-qc/1205.2586 (2012),
  \href{http://arxiv.org/abs/1205.2586 }{arXiv:gr-qc/1205.2586}.

\bibitem[Sto12e]{Sto11e}
\bysame, \emph{Schwarzschild singularity is semi-regularizable},
  \href{http://dx.doi.org/10.1140/epjp/i2012-12083-1}{Eur. Phys. J. Plus}
  \textbf{127} (2012), no.~83, 1--8,
  \href{http://arxiv.org/abs/1111.4837}{arXiv:gr-qc/1111.4837}.

\bibitem[Sto13a]{Sto13a}
C.~Stoica, \emph{{Singular General Relativity}}, Ph.D. Thesis (2013),
  \href{http://arxiv.org/abs/1301.2231}{arXiv:math.DG/1301.2231}.

\bibitem[Sto13b]{Sto11h}
O.~C. Stoica, \emph{{B}ig {B}ang singularity in the
  {F}riedmann-{L}ema{\^i}tre-{R}obertson-{W}alker spacetime}, The International
  Conference of Differential Geometry and Dynamical Systems (2013),
  \href{http://arxiv.org/abs/1112.4508}{arXiv:gr-qc/1112.4508}.

\bibitem[Sto13c]{Sto11g}
\bysame, \emph{{K}err-{N}ewman solutions with analytic singularity and no
  closed timelike curves}, To appear in U.P.B. Sci. Bull., Series A (2013),
  \href{http://arxiv.org/abs/1111.7082}{arXiv:gr-qc/1111.7082}.

\bibitem[Sto13d]{Sto12c}
\bysame, \emph{On the {W}eyl curvature hypothesis}, Ann. of Phys. \textbf{338}
  (2013), 186--194,
  \href{http://arxiv.org/abs/1203.3382}{arXiv:gr-qc/1203.3382}.

\bibitem[Sto14]{Sto12b}
\bysame, \emph{Einstein equation at singularities}, Central European Journal of
  Physics (2014), 1--9 (English),
  \href{http://arxiv.org/abs/1203.2140}{arXiv:gr-qc/1203.2140}.

\bibitem[Str95]{strominger95houches}
A.~Strominger, \emph{{L}es {H}ouches lectures on black holes}, arXiv preprint
  hep-th/9501071 (1995),
  \href{http://arxiv.org/abs/hep-th/9501071}{arXiv:hep-th/9501071}.

\bibitem[STU93]{susskind1993stretched}
Leonard Susskind, Larus Thorlacius, and John Uglum, \emph{The stretched horizon
  and black hole complementarity}, Physical Review D \textbf{48} (1993), no.~8,
  3743.

\bibitem[SV04]{singh2004quantum-bh}
TP~Singh and Cenalo Vaz, \emph{The quantum gravitational black hole is neither
  black nor white}, International Journal of Modern Physics D \textbf{13}
  (2004), no.~10, 2369--2373.

\bibitem[SV12]{visinescu2012bianchi}
B.~Saha and M.~Vi{\c{s}}inescu, \emph{Bianchi type-{I} string cosmological
  model in the presence of a magnetic field: classical versus loop quantum
  cosmology approaches}, Astrophysics and Space Science \textbf{339} (2012),
  no.~2, 371--377.

\bibitem[tHV74]{HV74qg}
G.~'t~Hooft and M.~Veltman, \emph{{One loop divergencies in the theory of
  gravitation}}, Annales de l'Institut Henri Poincar{\'e}: Section A, Physique
  th{\'e}orique \textbf{20} (1974), no.~1, 69--94.

\bibitem[Tod87]{Tod87}
K.~P. Tod, \emph{{Quasi-local Mass and Cosmological Singularities}}, Class.
  Quant. Grav. \textbf{4} (1987), 1457.

\bibitem[Tod90]{Tod90}
\bysame, \emph{{Isotropic Singularities and the $\gamma=2$ Equation of State}},
  Class. Quant. Grav. \textbf{7} (1990), L13--L16.

\bibitem[Tod91]{Tod91}
\bysame, \emph{{Isotropic Singularities and the Polytropic Equation of State}},
  Class. Quant. Grav. \textbf{8} (1991), L77.

\bibitem[Tod92]{Tod92}
\bysame, \emph{{Isotropic Singularities}}, Rend. Sem. Mat. Univ. Politec.
  Torino \textbf{50} (1992), 69--93.

\bibitem[Tod02]{Tod02}
\bysame, \emph{{Isotropic Cosmological Singularities}}, The Conformal Structure
  of Space-Time (2002), 123--134.

\bibitem[Tod03]{Tod03}
\bysame, \emph{{Isotropic Cosmological Singularities: Other Matter Models}},
  Class. Quant. Grav. \textbf{20} (2003), 521.

\bibitem[Ume10]{umetsu2010tunneling}
Koichiro Umetsu, \emph{Tunneling mechanism in kerr-newman black hole and
  dimensional reduction near the horizon}, Physics Letters B \textbf{692}
  (2010), no.~1, 61--63.

\bibitem[UO08]{udriste2008euler}
C.~Udri\c{s}te and D.~Opri\c{s}, \emph{{Euler-Lagrange-Hamilton dynamics with
  fractional action}}, WSEAS Transactions on Mathematics \textbf{7} (2008),
  no.~1, 19--30.

\bibitem[Vi{\c{s}}09]{visinescu2009bianchi}
M.~Vi{\c{s}}inescu, \emph{Bianchi type-{I} string cosmological model in the
  presence of a magnetic field: classical and quantum loop approach}, Romanian
  Reports in Physics \textbf{61} (2009), no.~3, 427--435.

\bibitem[Wal37]{WAL37}
A.~G. Walker, \emph{{On Milne's Theory of World-Structure}}, Proceedings of the
  London Mathematical Society \textbf{2} (1937), no.~1, 90.

\bibitem[Wal84]{Wal84}
R.~M. Wald, \emph{{General Relativity}}, University Of Chicago Press, June
  1984.

\bibitem[Wat12]{Watcharangkool2012}
Apimook Watcharangkool, \emph{The algebraic properties of black holes in higher
  dimension},
  \href{https://workspace.imperial.ac.uk/theoreticalphysics/Public/MSc/Dissertations/2012/Apimook
  Watcharangkool
  Dissertation.pdf}{https://workspace.imperial.ac.uk/theoreticalphysics/Public/MSc/Dissertations/2012/Apimook
  Watcharangkool Dissertation.pdf}.

\bibitem[Wei79]{Wein79AS}
S.~Weinberg, \emph{Ultraviolet divergences in quantum theories of
  gravitation.}, General relativity: an Einstein centenary survey, vol.~1,
  1979, pp.~790--831.

\end{thebibliography}
\newcommand{\etalchar}[1]{$^{#1}$}
\providecommand{\bysame}{\leavevmode\hbox to3em{\hrulefill}\thinspace}
\providecommand{\MR}{\relax\ifhmode\unskip\space\fi MR }
\providecommand{\MRhref}[2]{%
  \href{http://www.ams.org/mathscinet-getitem?mr=#1}{#2}
}
\providecommand{\href}[2]{#2}

\end{document}